\documentclass[fleqn,usenatbib]{mnras}

\usepackage[T1]{fontenc}
\usepackage{ae,aecompl}
\usepackage{graphicx}
\usepackage{amsmath}
\usepackage{newtxtext,newtxmath}

\title[Investigation of stellar activity using VAE]{Investigation of stellar magnetic activity using variational autoencoder based on low-resolution spectroscopic survey}
\author[Y.~Xiang, S.~Gu and D. ~Cao]{Yue Xiang,$^{1,2}$\thanks{E-mails: xy@ynao.ac.cn (YX); shenghonggu@ynao.ac.cn(SG)} Shenghong Gu,$^{1,2,3\star}$ and Dongtao Cao$^{1,2}$\\
$^{1}$Yunnan Observatories, Chinese Academy of Sciences, Kunming 650216, China\\
$^{2}$Key Laboratory for the Structure and Evolution of Celestial Objects, Chinese Academy of Sciences, Kunming 650216, China\\
$^{3}$School of Astronomy and Space Science, University of Chinese Academy of Sciences, Beijing 101408, China\\
}

\date{Accepted XXX. Received YYY; in original form ZZZ}

\pubyear{2022}

\begin{document}
\label{firstpage}
\pagerange{\pageref{firstpage}--\pageref{lastpage}}
\maketitle

\begin{abstract}
We apply the variational autoencoder (VAE) to the LAMOST-K2 low-resolution spectra to detect the magnetic activity of the stars in the K2 field. After the training on the spectra of the selected inactive stars, the VAE model can efficiently generate the synthetic reference templates needed by the spectral subtraction procedure, without knowing any stellar parameters. Then we detect the peculiar spectral features, such as chromospheric emissions, strong nebular emissions and lithium absorptions, in our sample. We measure the emissions of the chromospheric activity indicators, H$\alpha$ and Ca~{\sc ii} infrared triplet (IRT) lines, to quantify the stellar magnetic activity. The excess emissions of H$\alpha$ and Ca~{\sc ii} IRT lines of the active stars are correlated well to the rotational periods and the amplitudes of light curves derived from the K2 photometry. We degrade the LAMOST spectra to simulate the slitless spectra of the China Space Station Telescope (CSST) and apply the VAE to the simulated data. For cool active stars, we reveal a good agreement between the equivalent widths (EWs) of H$\alpha$ line derived from the spectra with two resolutions. The result indicates the ability of identifying the magnetically active stars in the future CSST survey, which will deliver an unprecedented large database of low-resolution spectra as well as simultaneous multi-band photometry of stars. 
\end{abstract}


\section{Introduction}

Cool stars with outer convective envelopes show solar-like active phenomena, such as starspots, plages and flares, which are driven by the stellar magnetic field. The current solar and stellar dynamo theories demonstrate that the magnetic field is generated beneath the convection zone and is related to the interaction between convection and rotation \citep{charbonneau2014}, and thus the strength of stellar magnetic field is dependent on the depth of the convection envelope and the rotational rate. 

To study the stellar magnetic activity, the activity-sensitive spectral lines, such as H$\alpha$, H$\beta$, Ca~{\sc ii} H \& K, He~{\sc i} D3 and Ca~{\sc ii} infrared triplet (IRT), are used as indicators. The spectral subtraction technique, which removes the inactive reference from the observed spectrum, is commonly used for measuring the chromospheric emissions \citep{gunn1997,montes1997,frasca2015}. Recent spectroscopic surveys, such as the Sloan Digital Sky Survey (SDSS; \citealt{york2000,abazajian2003}) and the Large Sky Area Multi-Object Fiber Spectroscopic Telescope (LAMOST, also named Guo Shoujing Telescope; \citealt{cui2012,deng2012}), have constructed a huge database of low-resolution stellar spectra, which is a treasure trove for the study of stellar activity. On the other hand, with the high-precision photometric survey, like Kepler/K2 \citep{koch2010}, many authors conducted vast researches on the stellar magnetic activity, such as chromospheric activity, starspots, differential rotation and activity cycles \citep{brandenburg2018,velloso2020}.

The China Space Station Telescope (CSST) is a planned 2-meter space survey telescope with field of view (FoV) of 1.1 deg$^{2}$, which will be equipped with multi-band imager and slitless spectrograph \citep{zhan2011,gong2019}. The slitless spectrograph has a resolution power of R $\geq200$ and a spectral coverage of 255-1000~nm. The limiting magnitude of the slitless spectroscopic observation is about 23 mag for the all-sky survey and is 24 mag for the deep field. The photometric and slitless spectroscopic observations will be taken by CSST simultaneously in one exposure, which will make it very helpful for the research on the stellar magnetic activity.

Machine learning is a powerful tool for the astrophysical data mining, which has been widely used for extracting information from massive photometric and spectroscopic data \citep{richards2011,neila2018}. Among all the techniques, the deep neural networks model currently achieves large success in various tasks of astronomy, such as detection, classification and identification (e.g. \citealt{dominguez2018,lanusse2018}). \citet{skoda2020} used the active deep learning to search for the emission-line stars.

In this work, we aim to build an unsupervised neural networks model which efficiently generates the template spectra for the spectral subtraction procedure, without knowing stellar parameters and then to characterize the stellar magnetic activity using the LAMOST-K2 low-resolution spectra. We also explore the possibility to detect the magnetically active stars with the planed CSST slitless spectroscopic survey. In Section 2, we describe the LAMOST-K2 data set and the simulated CSST slitless spectra. In Section 3, we introduce the details of the model architecture, the data preparation and model training. In Section 4, we show the reconstruction results and give the relative discussion. Finally, we summarise this work in Section 5.

\section{Data set}

LAMOST is a 4-meter reflecting Schmidt telescope located at the Xinglong station of National Astronomical Observatories of China. It has a large FoV of 5 degrees and is equipped with 4000 fibres at the focal plane. The LAMOST low-resolution spectrum has a resolution of 1800 at 5500 \AA\ and a wavelength range of 370--900~nm. The flux of the public LAMOST spectra are relatively calibrated \citep{song2012,xiang2015}.

The LAMOST-Kepler (LK) project aims to take advantage of the efficient spectroscopic survey of the LAMOST to observe the stars in the Kepler field \citep{fu2020}. The project has collected 227870 low-resolution spectra \citep{decat2015,zong2018}. Furthermore, the LAMOST-K2 (LK2) project, which targets the stars falling in the K2 field, has collected 160619 low-resolution spectra of 84012 stars \citep{wang2020}, which are included in the LAMOST Data Release 6 (DR6). Although their main purpose is the asteroseismic science, they shed light on the study of the stellar magnetic activity (e.g. \citealt{frasca2016,zhang2020, zhang2021})

\begin{figure}
\centering
\includegraphics[width=0.45\textwidth]{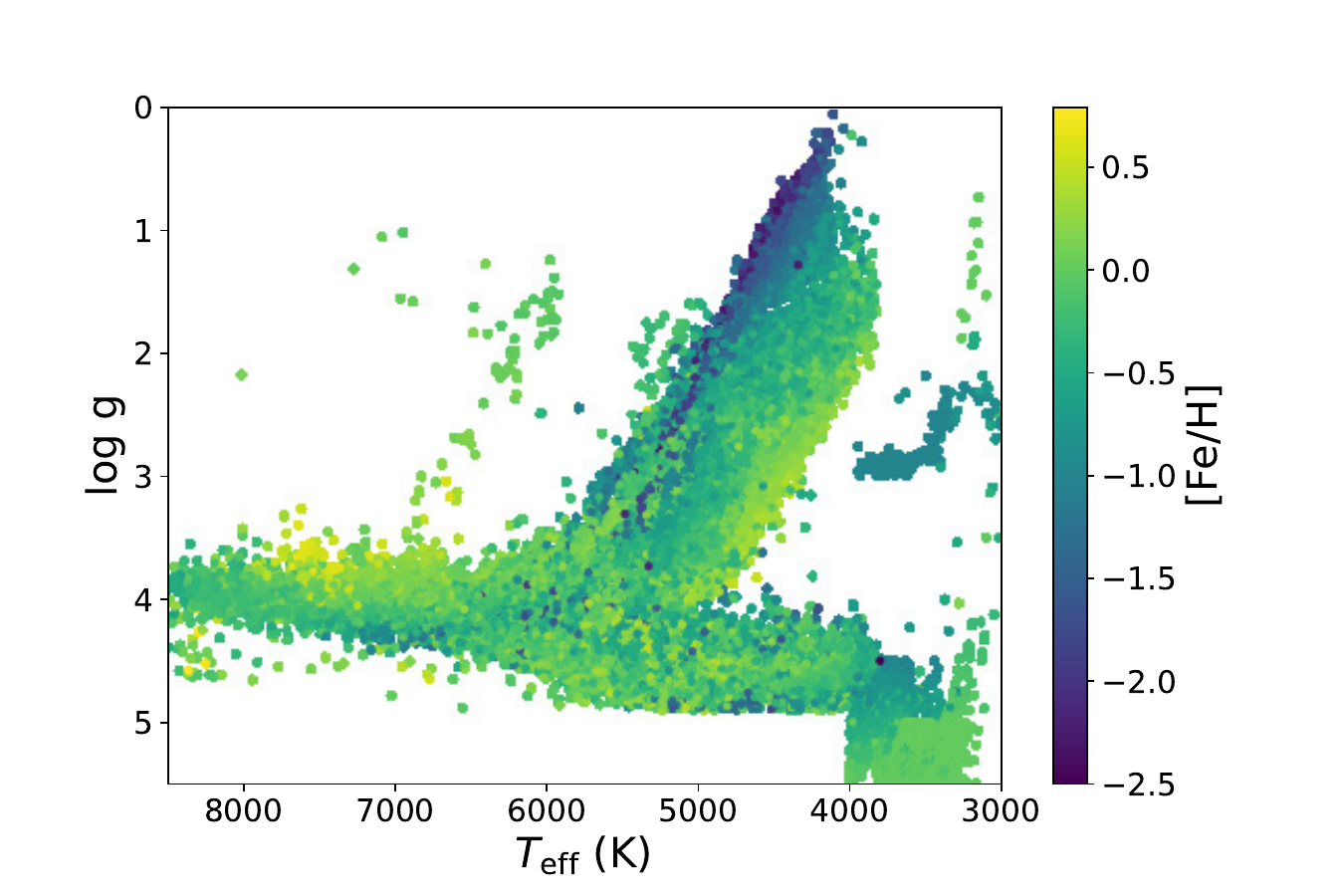}
\caption{Distribution of the stellar parameters of our sample. The adopted parameters are derived by the LASP.}
\label{fig:hr}
\end{figure}

We acquired the low-resolution stellar spectra in the LK2 survey project from the LAMOST website, according to the list of \citet{wang2020}. Two thresholds were set for the selection: the stellar effective temperature ($T_{\rm eff}$) should be in the range of 3000-8500 K and the signal-to-noise ratio (SNR) at the ${\it r}$ band should be larger than 10. For our purpose, the stars cooler than 3000 K suffer two problems: the number is too small and the active fraction is too high to select inactive stars \citep{west2004}. As a result of the selection, we obtained 132086 stellar spectra. Fig. \ref{fig:hr} shows the distribution of the stellar parameters, the $T_{\rm eff}$, the surface gravity ($\log g$) and the metal abundance ([Fe/H]), for our sample. The adopted stellar parameters are derived by the LAMOST stellar parameter pipeline (LASP; \citealt{luo2015}) and were also acquired from the LAMOST website. The radial velocity corrections were performed for all spectra in our sample. The radial velocity of A- to K-type stars is obtained from the LAMOST website, which is derived by the LASP. To derive the radial velocity of the M-type stars, we used the method described by \citet{fang2016}, who calculated the cross-correlation between the spectra of M-type stars and the template spectra of \citet{bochanski2007} to derive the spectral line shifts.

\begin{figure}
\centering
\includegraphics[width=0.45\textwidth]{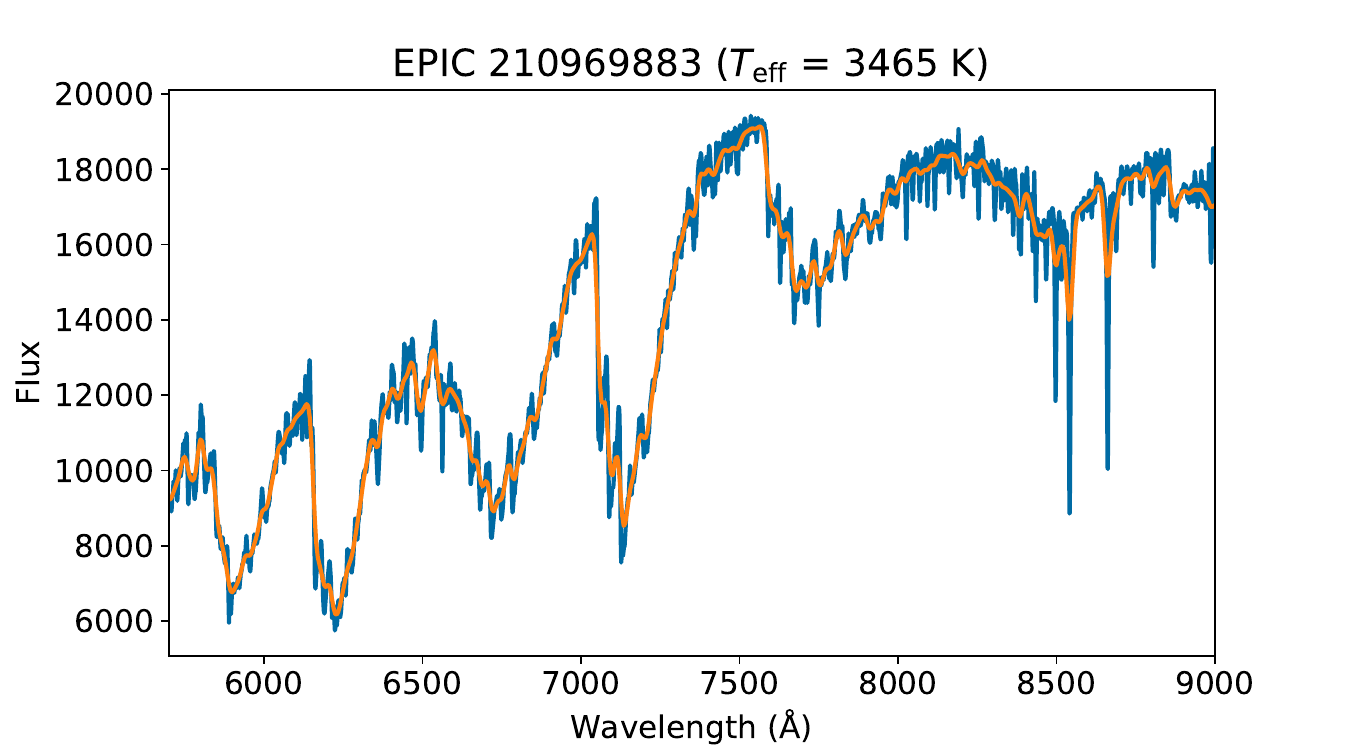}
\includegraphics[width=0.45\textwidth]{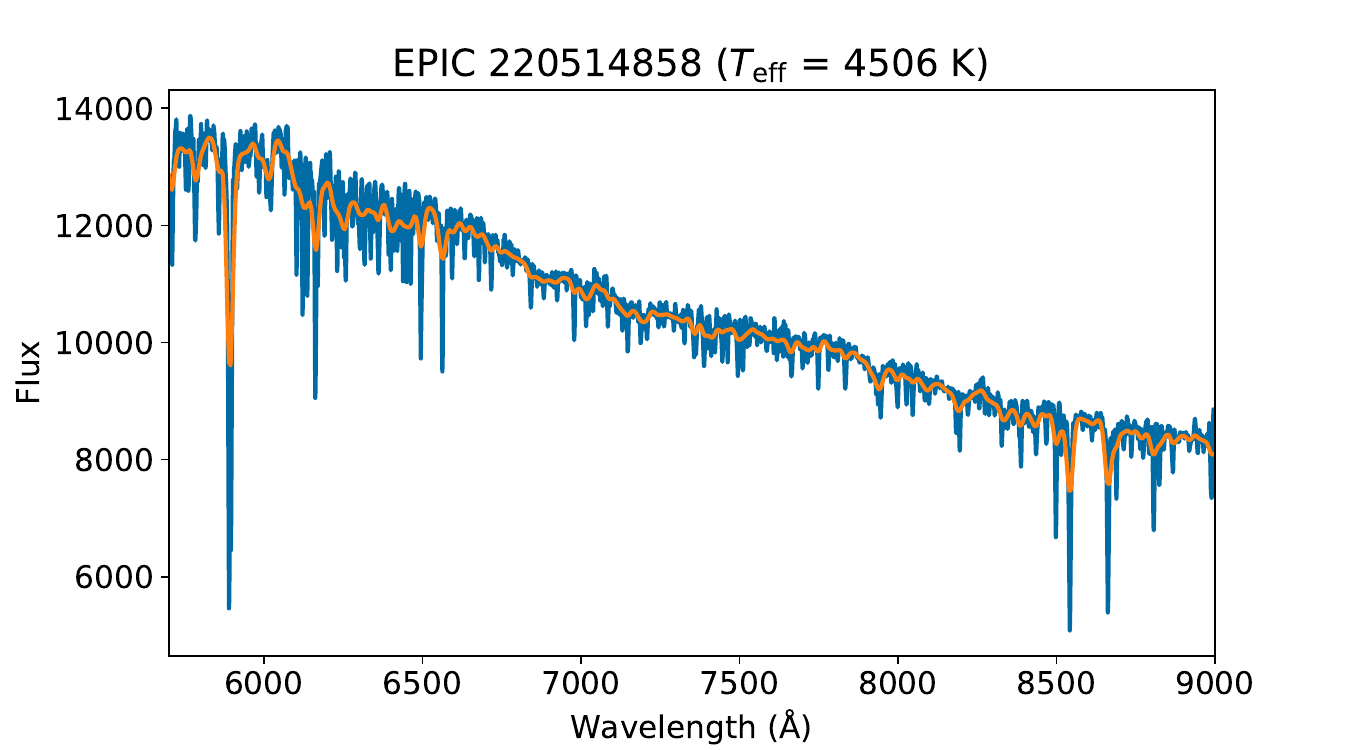}
\includegraphics[width=0.45\textwidth]{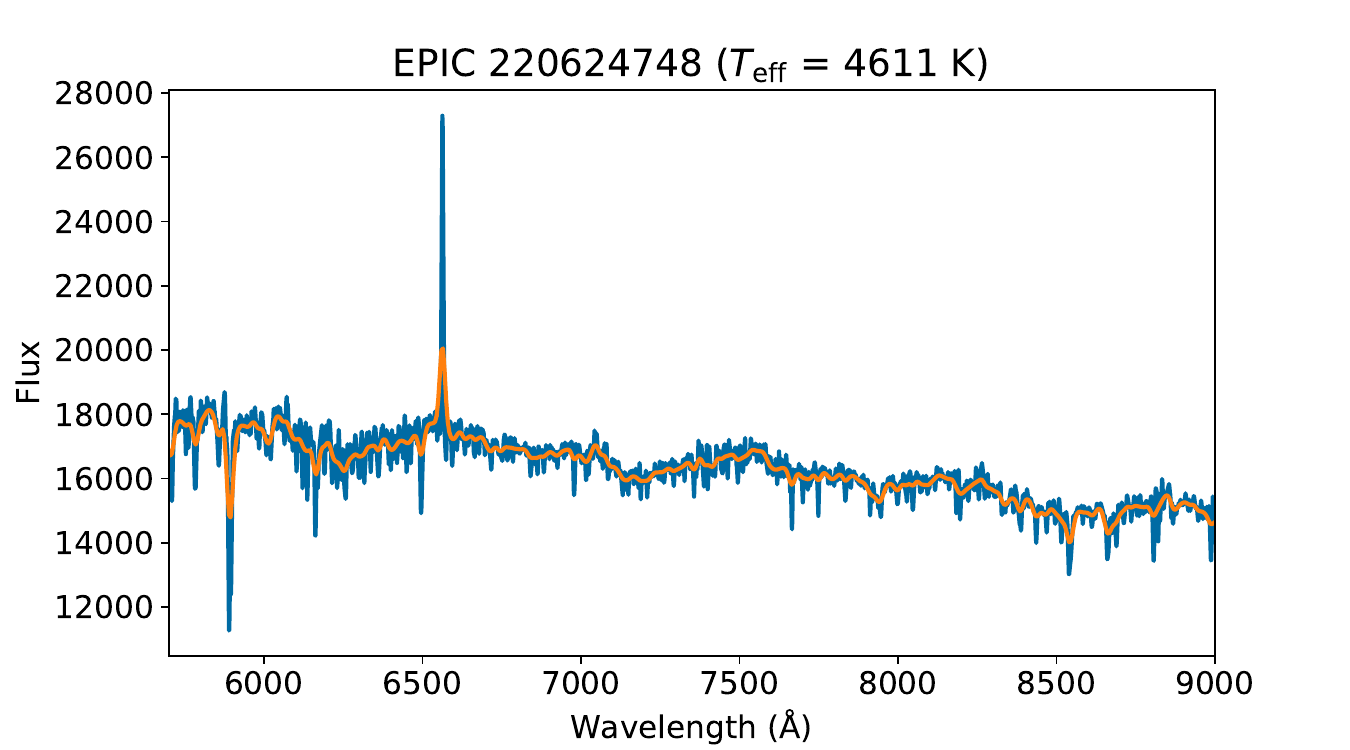}
\caption{Examples of the entire red band (5700--9000 \AA) LAMOST spectra (blue) and the corresponding degraded ones (orange).}
\label{fig:r250}
\end{figure}

To explore the ability of detecting active stars with the CSST slitless spectroscopic survey, we degraded the selected LAMOST spectra to simulate the CSST slitless spectra. The degradation was made by convolving the LAMOST spectra with a Gaussian profile. The degraded spectra has a resolution of about 250 at 5500 \AA, similar to the setup of the future CSST spectra. Hereafter we denote the original LAMOST spectra as the R1800 spectra and the degraded spectra as the R250 spectra for convenience. Fig. \ref{fig:r250} shows examples of the original red band (5700--9000 \AA) LAMOST spectra and the corresponding degraded ones. As can be seen, the degraded spectra preserve the emission feature of the H$\alpha$ line, though the peak becomes wider and much smaller. Hence the H$\alpha$ line is still a relatively prominent chromospheric indicator in the very-low-resolution spectra.

The R1800 spectra were interpolated onto a same grid with a linear wavelength increment of 1 \AA, since the neural networks model requires a uniform dimensionality of the input. The R250 spectra were interpolated onto a linear grid with an increment of 5 \AA. We normalized the continuum of the R1800 spectra with a Legendre polynomial. Whereas, the R250 spectra were normalized in a different way: divided by the median value of the fluxes within the wavelength range of 6500-6550 \AA. 

In this work, we did not use the entire wavelength range of the R1800 and R250 spectra. We extracted two bands that cover the regions of H$\alpha$ and Ca~{\sc ii} IRT, respectively, from the normalized R1800 spectra, which are 6370--6770 \AA\ and 8380--8780 \AA. We did not use Ca~{\sc ii} H \& K lines since the SNR at the blue end of the LAMOST spectra is much lower, especially for the M-type stars. The excess emission of Ca~{\sc ii} IRT is well correlated to that of Ca~{\sc ii} H \& K, and is also the important indicators of the stellar magnetic activity \citep{busa2007,martin2017}. For R250 spectra, we extracted the normalized flux at the wavelength range of 6000-7500 \AA, to analysis the emission of H$\alpha$, which is wide enough to be seen in the slitless spectra.

The different treatments on the R1800 and R250 spectra are because of the different spectral resolution and the corresponding analysis methods. For the R1800 spectra, we performed the spectral subtraction technique, which commonly requires the continuum-normalized spectra \citep{montes1997,frasca2016}, to analyse the residual emissions of the chromospheric indicators,  H$\alpha$ and Ca~{\sc ii} IRT. In the degraded spectra, the H$\alpha$ line is still a relatively prominent feature (Fig. \ref{fig:r250}), and we thus only analysed the H$\alpha$ line in R250 spectra. The normalization method for the R250 spectra preserved the original spectral shape, and the resulting spectra can be similar to the flux-calibrated CSST spectra with the same normalization.

\section{Methods}

In order to detect the interesting features in stellar spectra, we need to know the normal spectra firstly. One can generate the reference spectrum by observing a standard star or using a stellar model atmosphere. However, it either requires the accurate stellar parameters or consumes much time on the template matching through cross-correlation or least-square fitting, and the result is very dependent on the completeness and quality of the template spectra grid. In this section, we introduce an unsupervised data-driven model to generate the template spectra and describe the data preparation and model training.

\subsection{Variational Autoencoder}

The autoencoder (AE; \citealt{hinton2006}) is an hourglass shaped neural networks, which mainly consists of two components, encoder and decoder. The encoder learns to extract the latent features which describe the input data, and the decoder tries to reconstruct the data from the features extracted by the encoder. As usual, the dimensionality of the latent features is much less than that of the input dataset. Thus the AE is often used for dimensionality reduction \citep{hinton2006,yang2015}. Due to the nonlinearity, the AE performs much better at modelling spectra than the principle components analysis (PCA), which makes linear transformation \citep{portillo2020}. Meanwhile, the AE is often used for the anomaly detection.

AE is easy to overfit, especially when the dimensionality of the latent codes is large. In principle, if the dimension of the latent codes is same as that of the input data, the AE will learn to copy the original input to perform a perfect but useless reconstruction. A solution is the so-called denoising autoencoder (DAE), which is trained on a manually corrupted dataset and computes the loss function between the output and the original data.

The variational autoencoder (VAE; \citealt{kingma2014}) is a variant of AE, and is widely used in astronomy (e.g. \citealt{portillo2020,sedaghat2021}). Although the architecture of VAE is similar to that of AE, the concept behind is very different, and VAE is based on the variational Bayesian inference \citep{kingma2014}. The encoder component of VAE maps the input data into a distribution, rather than a fixed code, and the advantage is that the latent space becomes continuous. Thus VAE has a different architecture of the bottleneck, which produces the mean ($\mu$) and the logarithmic variance ($\log \sigma^{2}$) of the latent distribution, and then samples the latent variable ($z$) from the distribution. The VAE additionally introduces a regularization on the latent space, which is the Kullback-Leibler (KL) divergence between the generated latent distribution and a prior distribution. The loss function of the VAE is the evidence low-boundary (ELBO):
\begin{equation}
L_{\rm VAE} = L_{\rm recon}(x, x') + D_{\rm KL}(q_{\phi}(z|x) \| p_{\theta}(z))
\end{equation}
where the first term is the reconstruction loss for which we used the mean squared error (MSE). The second term is the KL divergence, which measures the difference between the latent distribution generated by the encoder ($q_{\phi}(z|x)$) and the prior distribution ($p_{\theta}(z)$), in which $\theta$ and $\phi$ are learnable parameters. In practice, the standard Gaussian distribution is usually chosen as the prior distribution.

\begin{figure*}
\centering
\includegraphics[width=0.6\textwidth]{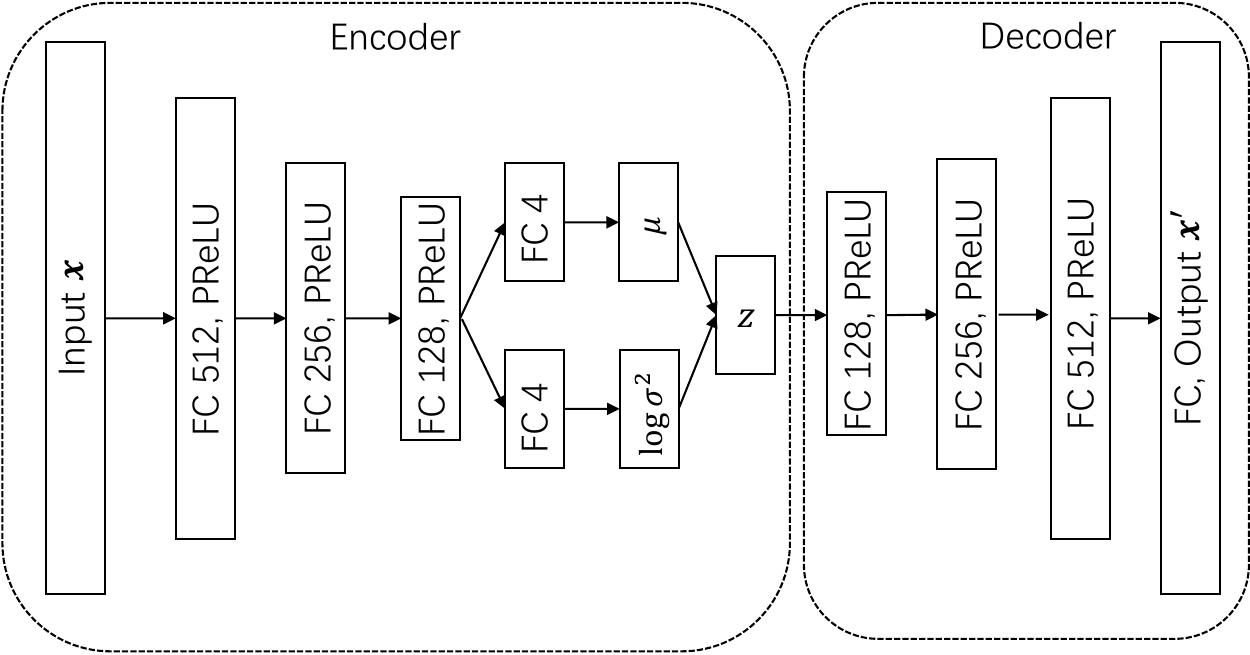}
\caption{Schematic diagram of the architecture of our VAE model.}
\label{fig:vae}
\end{figure*}

We built a VAE model using PyTorch \citep{paszke2019}, a deep learning framework in Python. The encoder and decoder of the model consisted of the fully-connected (FC) neural networks. In Fig. \ref{fig:vae}, we display the schematic diagram of the architecture of our VAE model, including the type and dimension of layers. For the R1800 and R250 spectra, the architecture of the VAE were the same, except for the input dimension (800 for R1800 and 300 for R250). The numbers of hidden layers were arbitrarily selected. Similar structures of fully connected VAEs can be found in many papers (e.g. \citealt{ichinohe2019,iwasaki2019,bastien2021}) for different purposes. We used the activation function of the Parametric Rectified Linear Units (PReLU; \citealt{he2015}) as the equation below,
\begin{equation}
{\rm PReLU}(x) =
\begin{cases}
x,     & \text{if x $\geq$ 0}\\
ax, & \text{if x<0}
\end{cases}
\end{equation}
where $a$ is a learnable parameter, which modifies the negative slope of the activation function. As a variant of Rectified Linear Units (ReLU), PReLU is better than ReLU at dealing with the all negative inputs.

\subsection{Preprocess and training}

\begin{figure}
\centering
\includegraphics[width=0.45\textwidth]{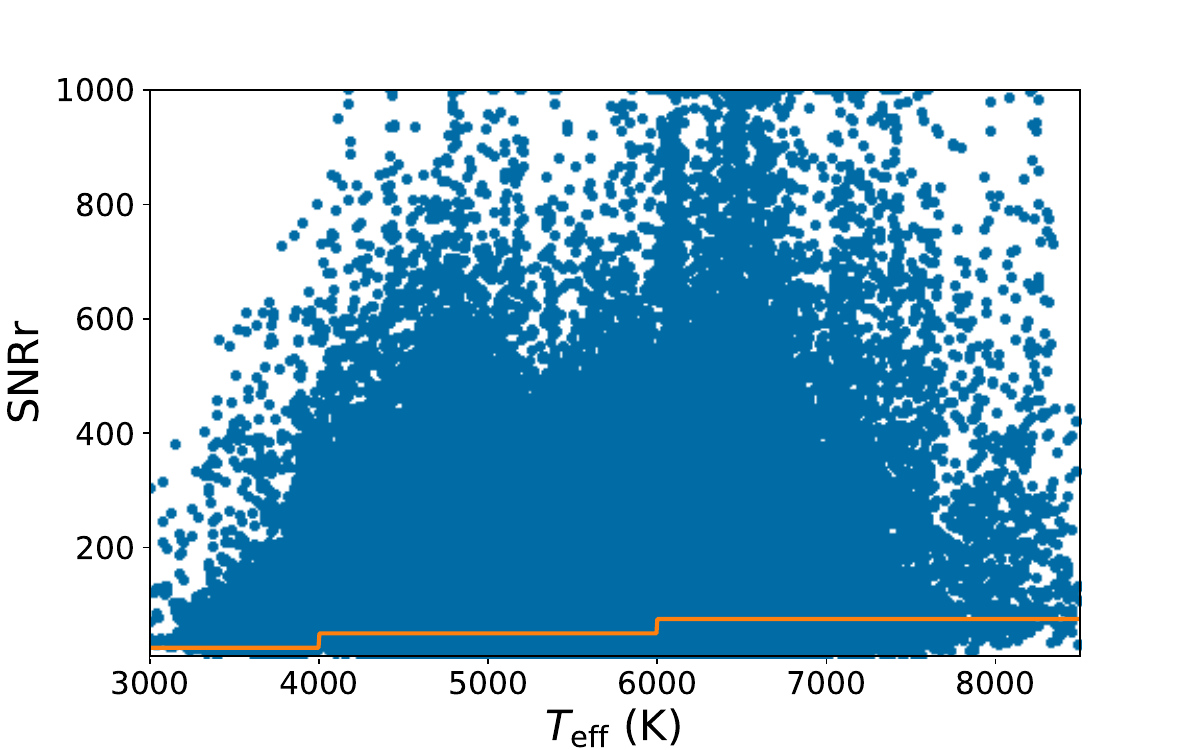}
\includegraphics[width=0.45\textwidth]{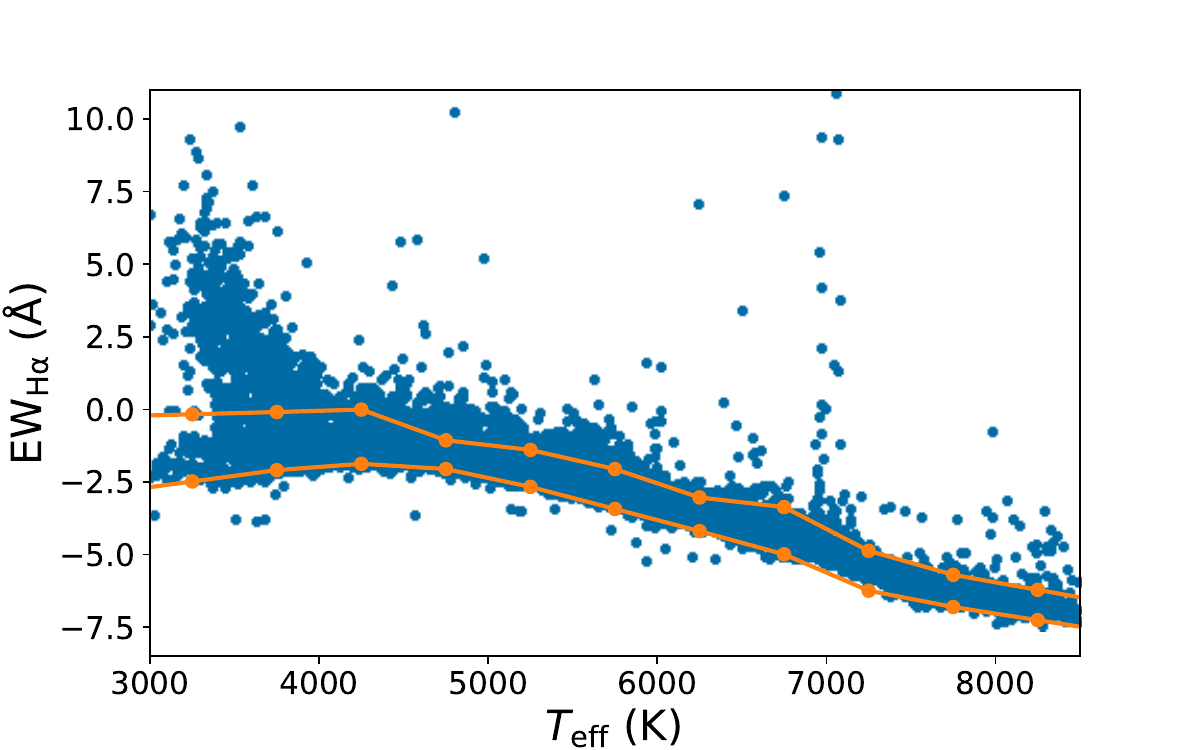}
\caption{Sample selection at the first step. The upper panels shows the distribution of SNRr vs the $T_{\rm eff}$, and the orange line denotes the lower boundary of the selection. The lower panel shows the EWs of the H$\alpha$ line vs the $T_{\rm eff}$. A positive EW represents the emission. The orange points show the upper and lower boundary of each $T_{\rm eff}$, and the line is the linear interpolation of these points.}
\label{fig:r1}
\end{figure}

\begin{figure}
\centering
\includegraphics[width=0.4\textwidth]{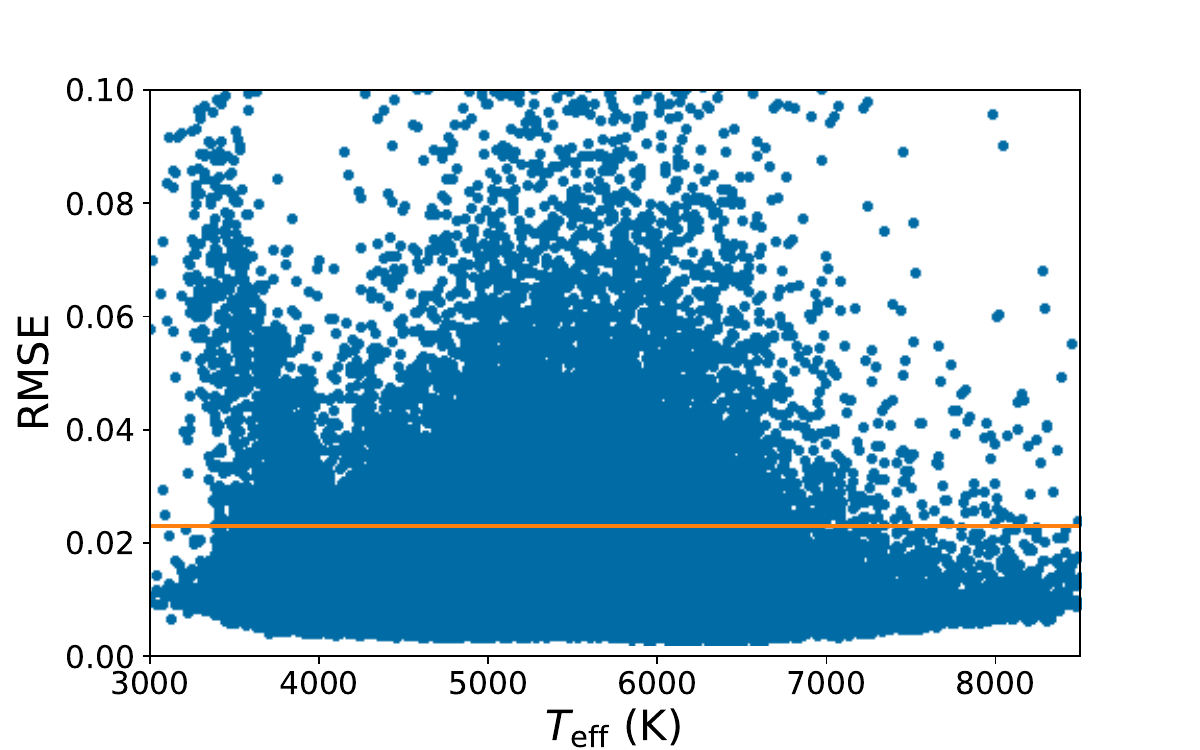}
\includegraphics[width=0.4\textwidth]{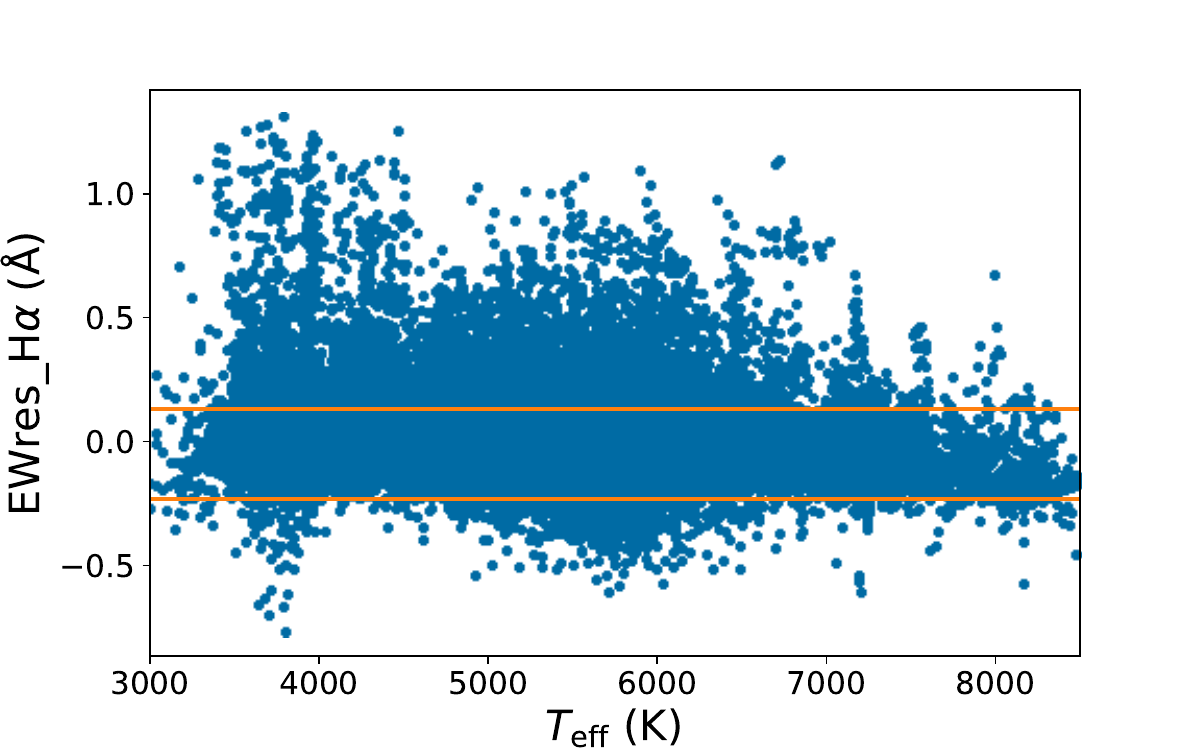}
\includegraphics[width=0.4\textwidth]{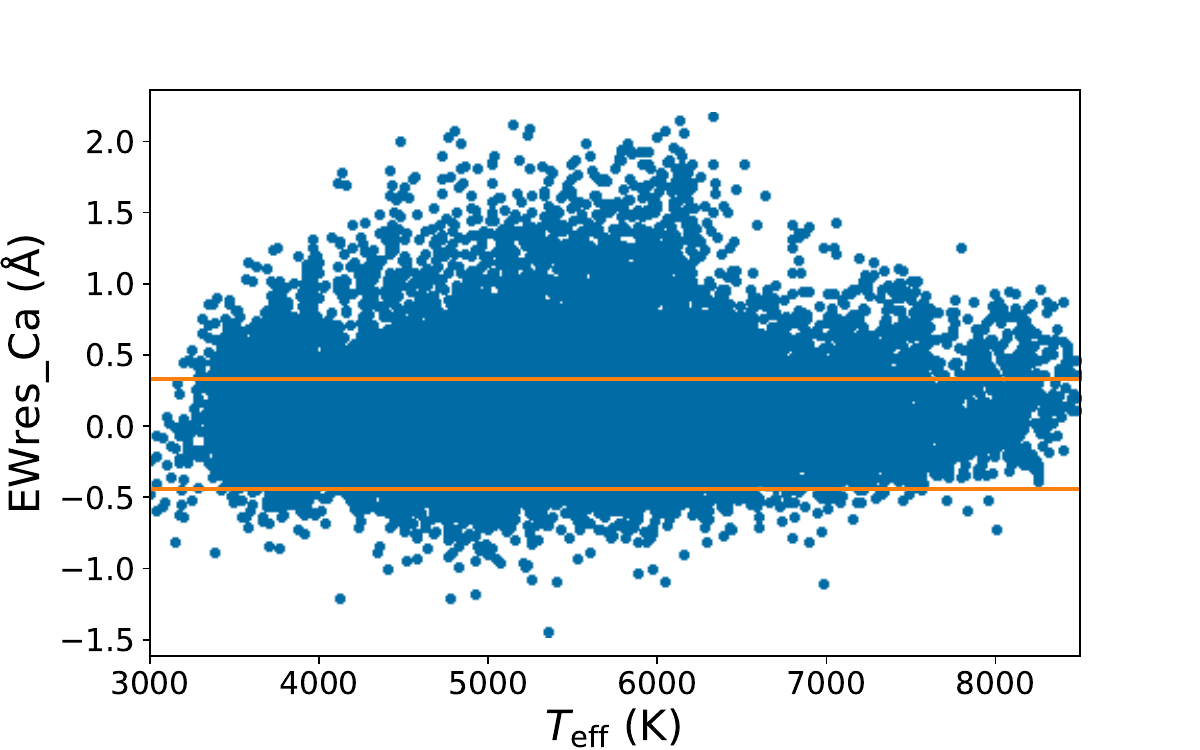}
\caption{Sample selection at the second step. The top panel shows the RMSE of the fits of the initial model. The middle and bottom panel show the residual EWs of the H$\alpha$ line and the summed value of the Ca~{\sc ii} IRT lines, respectively. The orange lines represent the boundary of the selection.}
\label{fig:r2}
\end{figure}

Before the model training, the preprocess including data selection and standardization is essential. Since we want to generate the inactive template spectra using the VAE model, we need to select a set of the inactive stellar spectra to construct the training dataset. We implemented a two-step procedure to select the normal R1800 spectra. Firstly, we set a criterion on the SNR at the ${\it r}$ band to select high-SNR spectra for the training. The lower limit of SNR is 25 for $T_{\rm eff}$ in the range of 3000-4000 K, 50 for those in the range of 4000-6000 K, and 75 for those in the range of 6000-8500 K. Then we simply derived the EWs of H$\alpha$ line by integrating over a wavelength interval of 20 \AA\ around 6563 \AA. The resulting EWs are shown in Fig. \ref{fig:r1}, where a positive EW means the emission. Since the EWs of H$\alpha$ line are much dependent on the $T_{\rm eff}$ of stars, it needs to take the $T_{\rm eff}$ into account to set the threshold of EWs. We binned the EWs on the $T_{\rm eff}$ with a width of 500 K. To remove active stars and outliers, we set the upper and lower boundary as 99th and 1st percentile of EWs in each $T_{\rm eff}$ bin hotter than 4000 K, and 60th and 1st percentile of EWs in each cooler than 4000 K, since the fraction of active M stars is much larger than hotter ones \citep{west2004}. Then we used the linear interpolation of these points as the boundary of the EWs. As a result, a total of 107604 spectra were selected (Fig. \ref{fig:r1}). After an initial training, we modelled the R1800 spectra. We derived the root-mean-square error (RMSE) and the residual EWs of the H$\alpha$ and Ca~{\sc ii} IRT lines from the residual spectra, as shown in Fig. \ref{fig:r2}. The upper boundary of RMSE is 0.022, which is 90th percentile of RMSE. The upper and lower thresholds of the residual EWs were set to 90th and 1st percentile, respectively, as shown in Fig. \ref{fig:r2}, in order to select the most inactive stars for the training sample. The number of the spectra in the final sample is 85832. These inactive samples were also used for the R250 model training.

Data standardization is essential for the machine learning algorithms, and it can improve the performance of the model. Without the data standardization, the dimension of the input with the largest variance may dominate the result. We used the standard scaler in the Scikit-learn \citep{pedregosa2011}, a machine learning package in Python, to fit the selected normalized spectra. The standard scaler removes the mean value from each dimension and rescales each dimension to unit variance. Since the standardized data has no physical meaning, it was purely used for the model training and inference.

In order to make the VAE model more robust to the abnormal features, such as cosmic rays and strong H$\alpha$ emission lines, we randomly added emission-like features and Gaussian noise to the input of the training dataset, similar to that for DAE.

\begin{figure}
\centering
\includegraphics[width=0.45\textwidth]{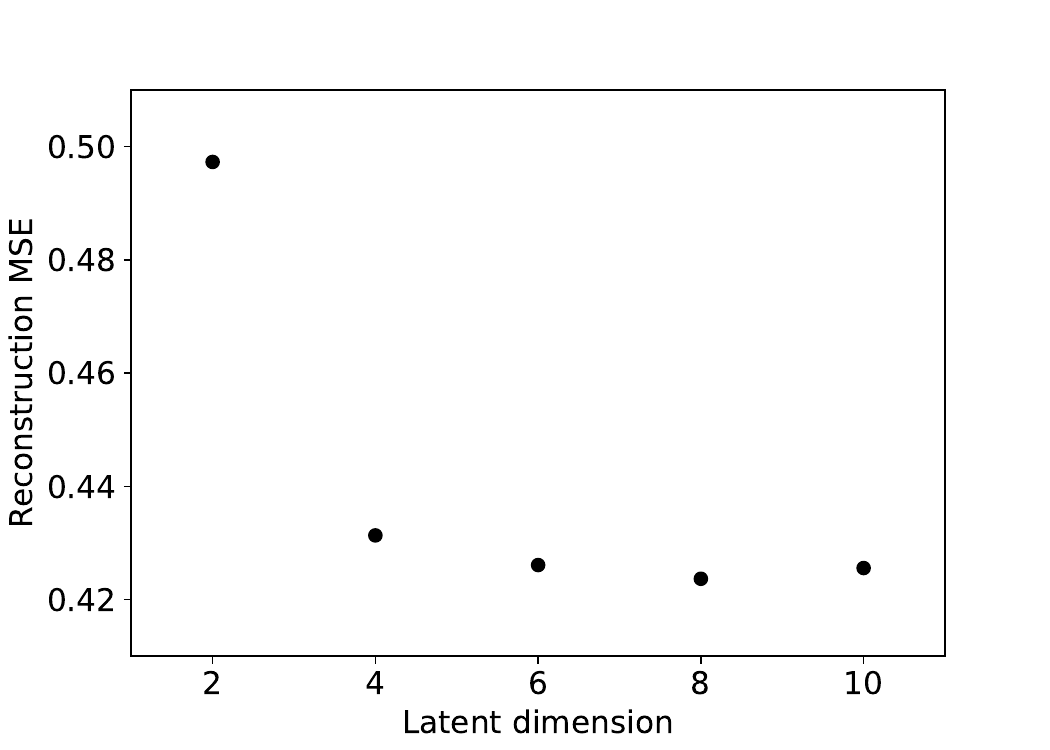}
\caption{The reconstruction MSE of the valid dataset as a function the latent dimension.}
\label{fig:ld}
\end{figure}

During the model training, we implemented the early stopping technique to prevent the neural networks model from overfitting. The selected inactive spectral dataset were randomly split into two subsets, the training and valid datasets, which are 80 and 20 per cent of the total number, respectively. At each training epoch, the early stopping method used the loss curve of the valid set to check if the model began to over fit and stopped the training process when the valid loss began to increase. We tried different numbers of the dimension of the latent space. Fig. \ref{fig:ld} shows the reconstruction MSE of the valid dataset versus different latent dimensions. We found that the reconstruction with the latent dimension of 4 is already sufficient for our spectral analysis.

\section{Results and discussion}

\subsection{Reconstruction}

\begin{figure}
\centering
\includegraphics[width=0.45\textwidth]{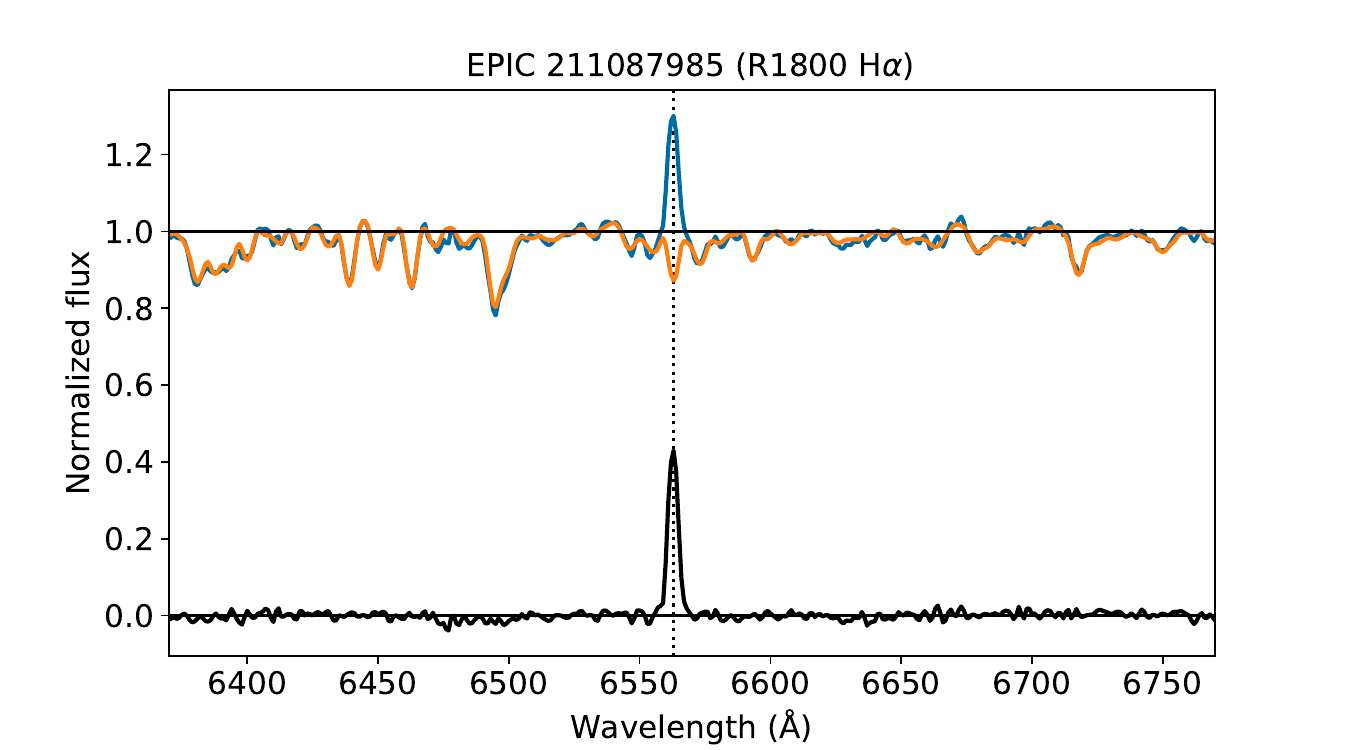}
\includegraphics[width=0.45\textwidth]{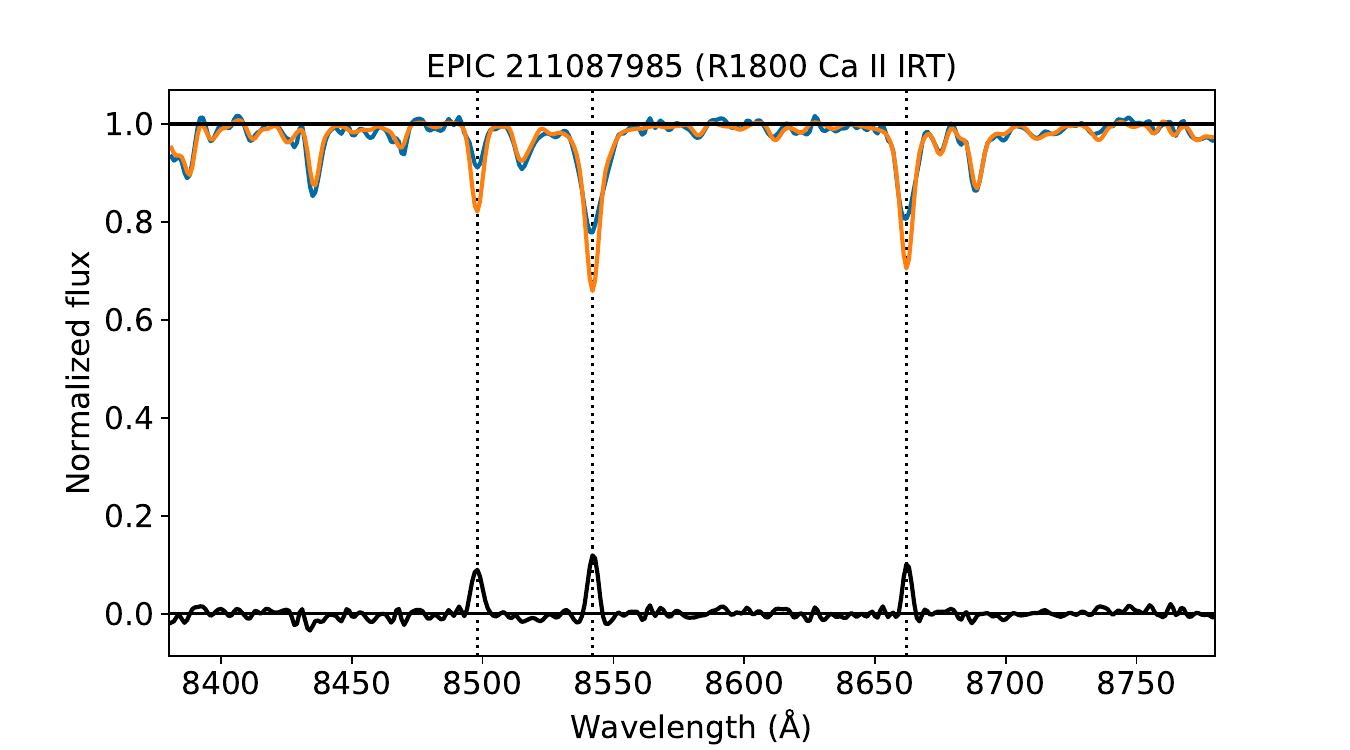}
\includegraphics[width=0.45\textwidth]{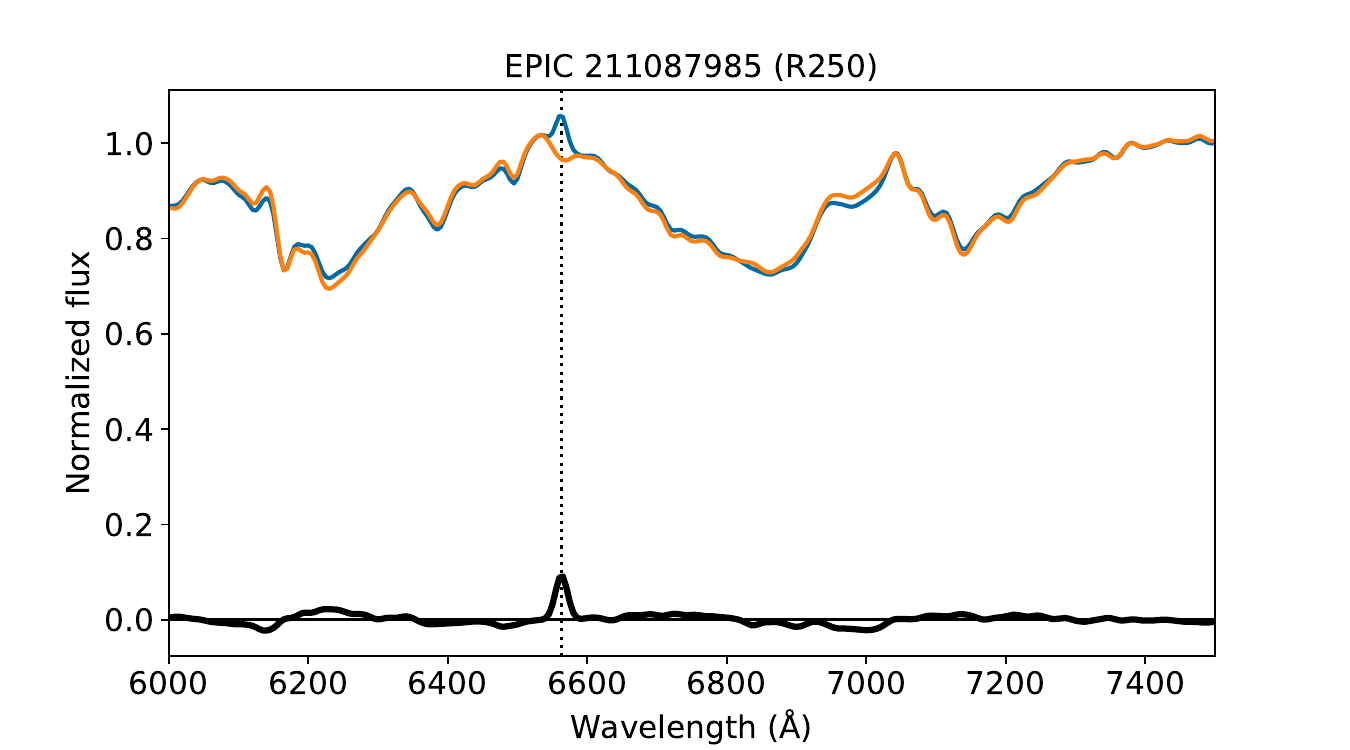}
\caption{Examples of the fits (orange) to the R1800 (top and middle) and R250 (bottom) normalized spectra of an active star (EPIC 211087985). H$\alpha$ and Ca~{\sc ii} IRT line centres are denoted by the vertical dotted lines. The residuals are shown in the bottom of each panel. Note that the normalization methods are different for the R1800 and R250 spectra.}
\label{fig:example}
\end{figure}

We separately used the VAE to model R1800 and R250 spectra, in an unsupervised way. Given an input of stellar spectrum, the VAE model can reconstruct the corresponding inactive template spectrum. Examples of fits to the two bands of the R1800 spectra of an active star, EPIC 211087985, are shown in the upper two panels of Fig. \ref{fig:example}. We performed a correction on the R250 spectra using the method in \citet{cotar2021}. We calculated the flux ratios between the R250 spectra and the VAE reconstructions ($R = F_{\rm R250}/F_{\rm VAE}$). Then the ratios were fitted by a low-order polynomial. At last we divided the R250 spectra by the fitted polynomial. The example of reconstruction of VAE for the R250 spectra of the active star EPIC 211087985 is shown in the bottom panel of Fig. \ref{fig:example}.

\begin{figure}
\centering
\includegraphics[width=0.45\textwidth]{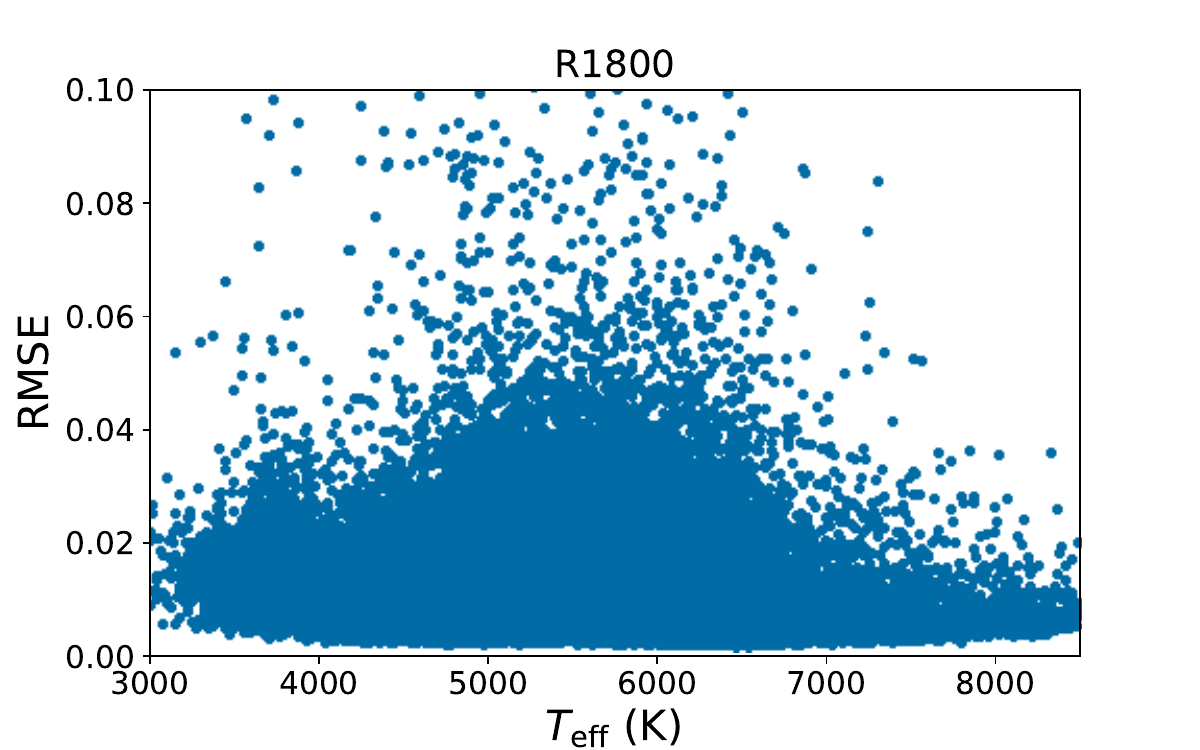}
\includegraphics[width=0.45\textwidth]{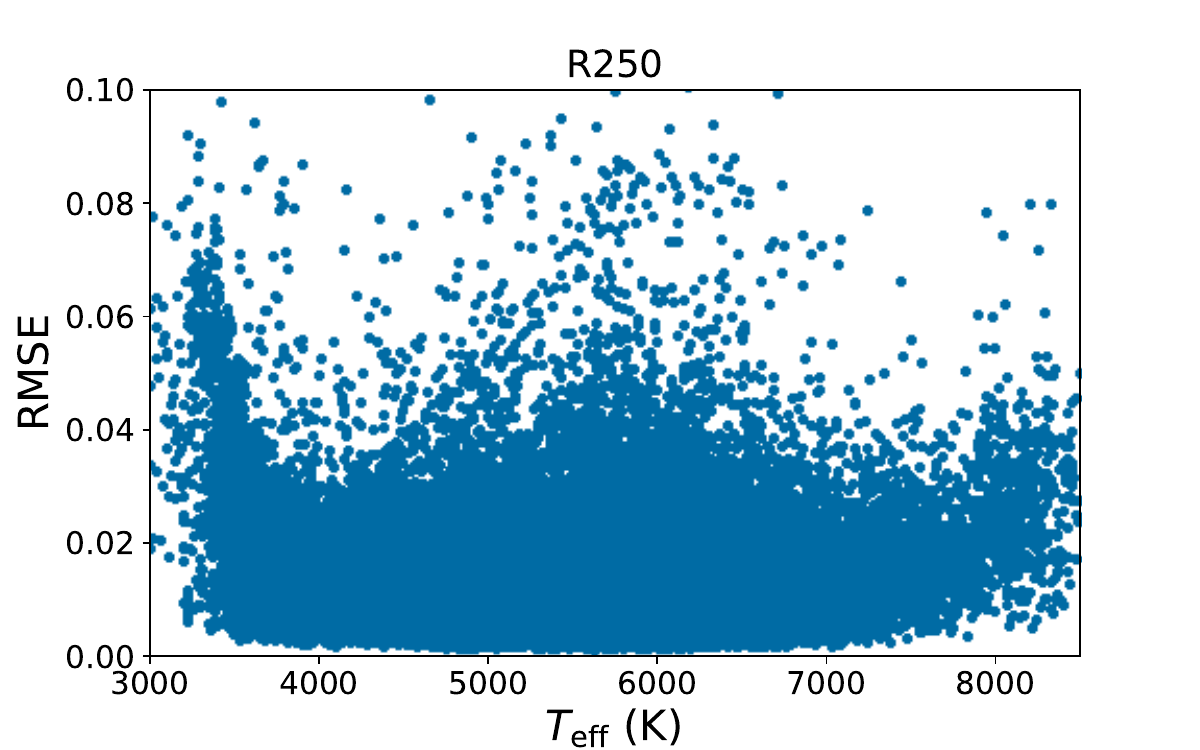}
\caption{The RMSE of the fits to the R1800 and R250 normalized spectra for different $T_{\rm eff}$. }
\label{fig:rmse}
\end{figure}

To estimate the performance of the VAE models, we derive the RMSE of the residual spectra outside the chromospheric active lines. Fig. \ref{fig:rmse} shows the distribution of the RMSE across different $T_{\rm eff}$. The overall RMSE of the fits to the R1800 and R250 spectra are 0.0157 and 0.0151, respectively. The performance of the R250 model on the M stars is worse than that for hotter ones, probably due to the relatively lower SNR and the higher inaccuracy in the flux calibrations, which leads to the errors in the spectral shape.

\begin{figure}
\centering
\includegraphics[width=0.45\textwidth]{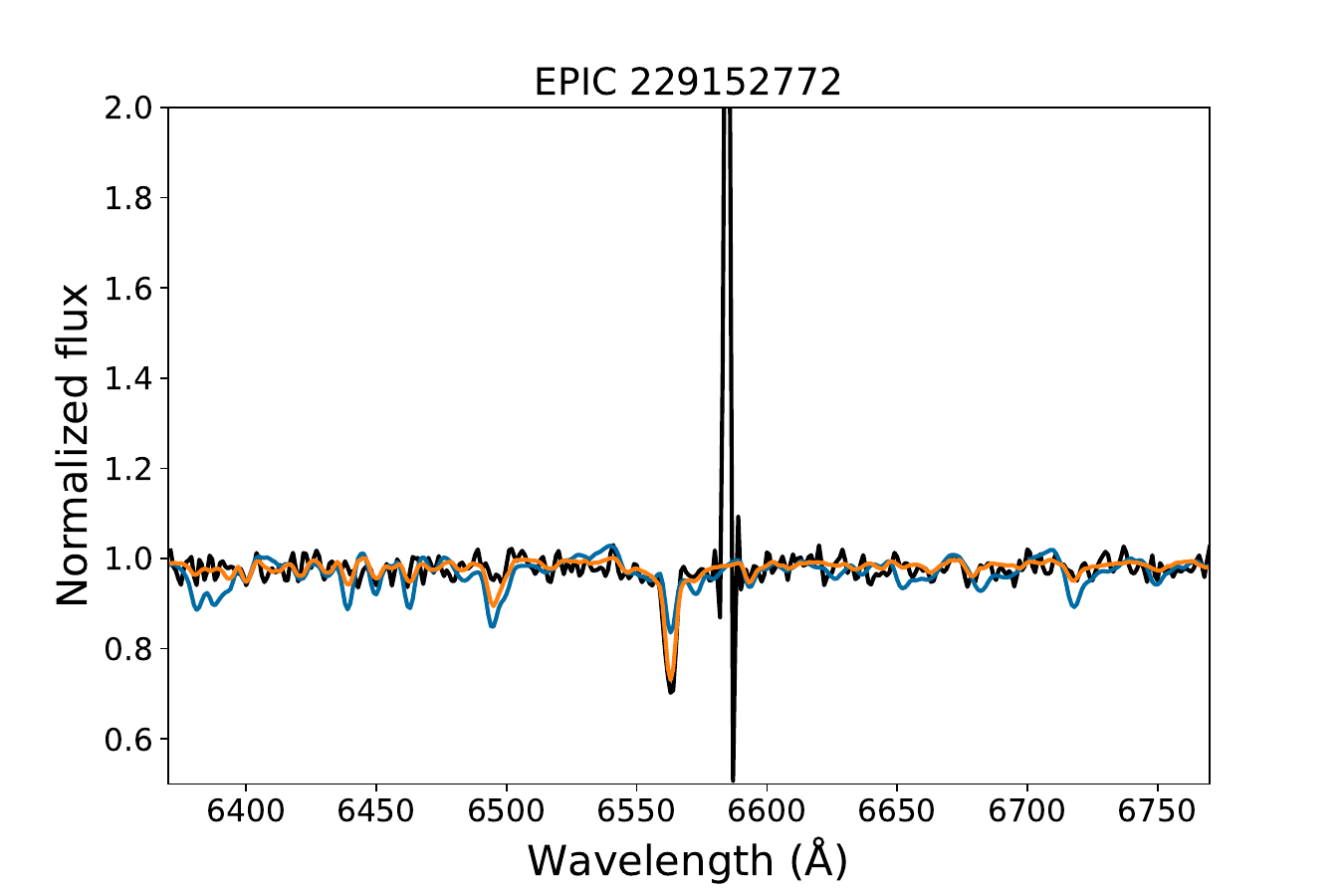}
\caption{The fits to the spectrum with a strong cosmic ray. The model trained on the data with manually added noise (orange) is more robust to the abnormal spectrum than that trained on the original data (blue).}
\label{fig:ef}
\end{figure}

After adding noise and emission features into the training set, our model is more robust to the strong emission-like features, such as the H$\alpha$, nebular forbidden emission lines and cosmic rays. Fig. \ref{fig:ef} shows an example of the fits to a spectrum with strong cosmic ray. The model which is trained by a noisy data performs better and can generate the proper template.

\begin{figure}
\centering
\includegraphics[width=0.45\textwidth]{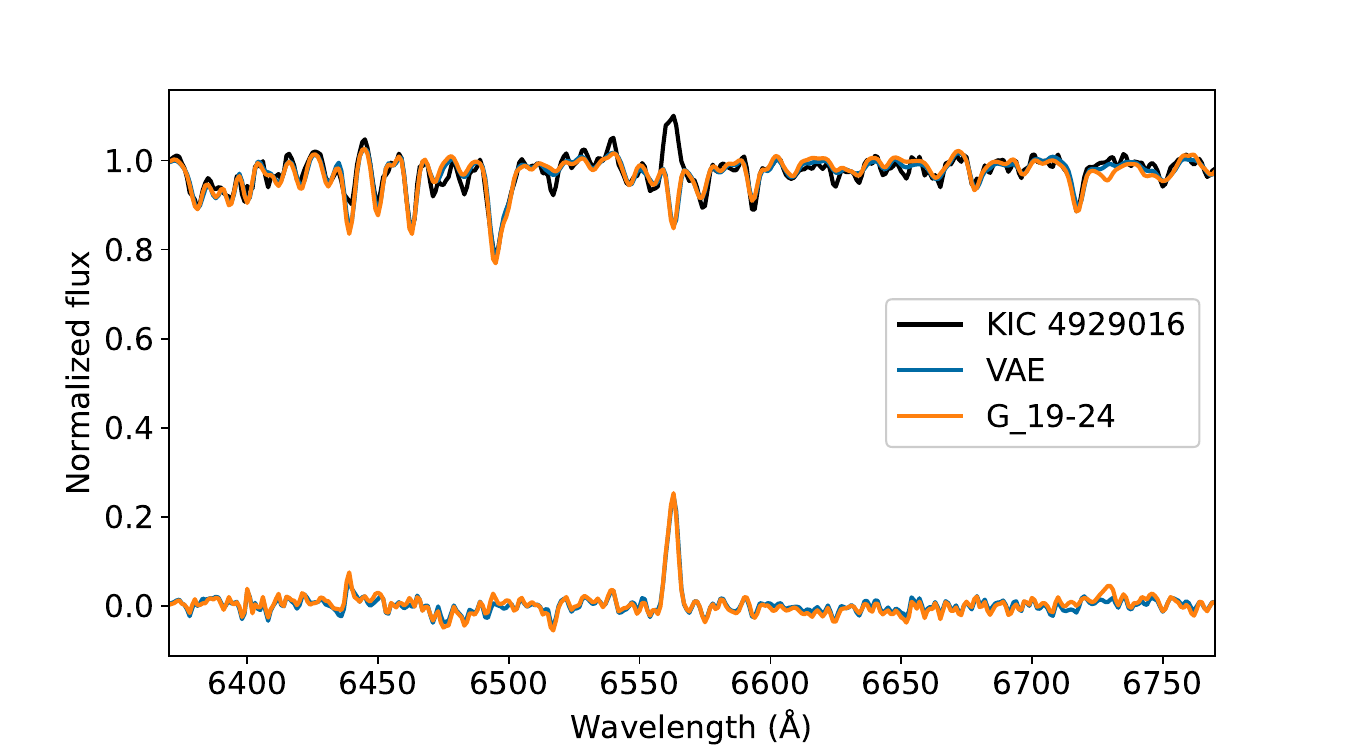}
\includegraphics[width=0.45\textwidth]{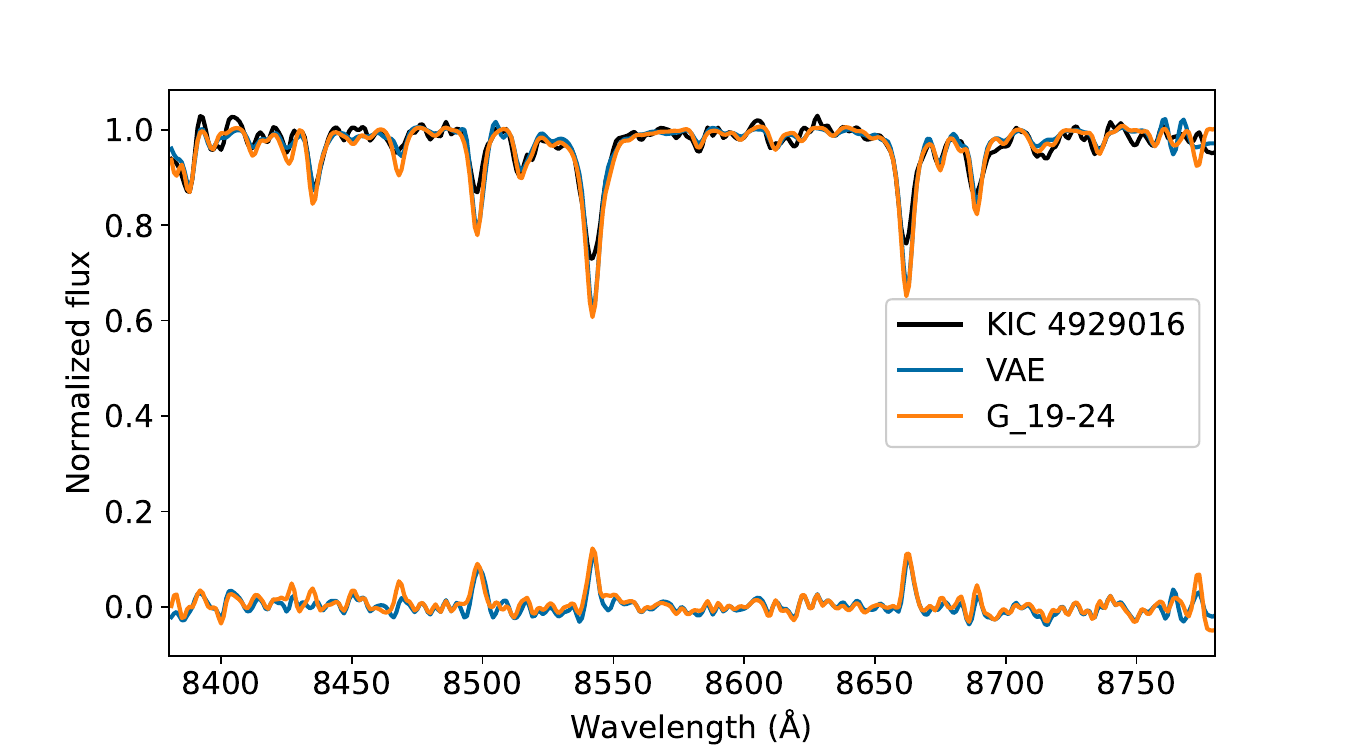}
\caption{The fits of VAE (blue) to the spectra of an active star, KIC4929016 (black) as well as the rotationally broadened spectra of the best-fit star (orange) from the CFLIB.}
\label{fig:kepler}
\end{figure}

The generalization of the model is important, which represents the ability of the model to handle a previously unseen new data. To demonstrate the generalization ability of our VAE model trained on the LK2 spectral data, we applied the trained model to the low-resolution stellar spectra database of the LK project. \citet{frasca2016} used the template spectra from the Indo-US coude feed stellar spectral library (CFLIB; \citealt{valdes2004}) to perform the template matching and spectral subtraction to estimate the stellar parameters and analyse the chromospheric activity with the LK data. Fig. \ref{fig:kepler} shows the spectra of KIC 4929016, the VAE reconstruction as well as the template spectrum, where the best-fit template star are provided in \citet{frasca2016}. It shows that the VAE model generates proper template spectrum without knowing stellar parameters and performs well on the new data.

\subsection{Anomaly detection}

\begin{figure}
\centering
\includegraphics[width=0.45\textwidth]{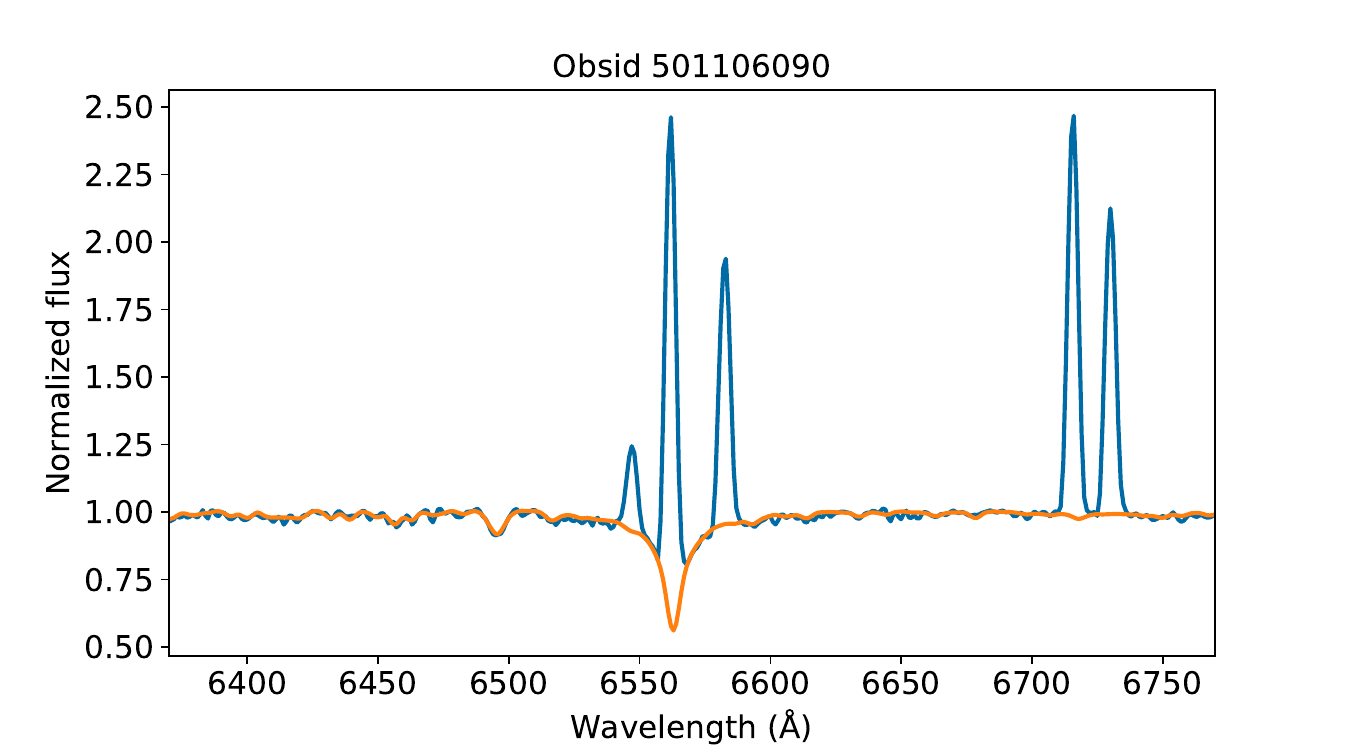}
\includegraphics[width=0.45\textwidth]{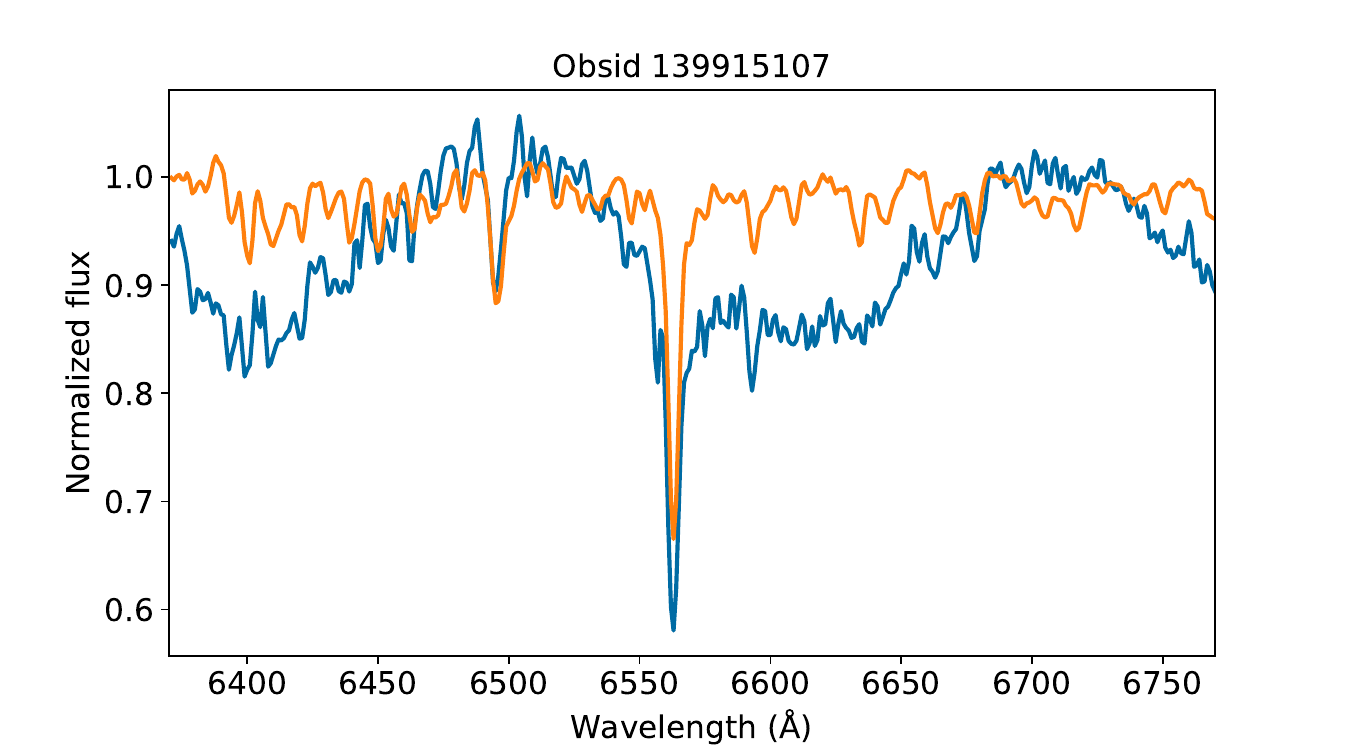}
\includegraphics[width=0.45\textwidth]{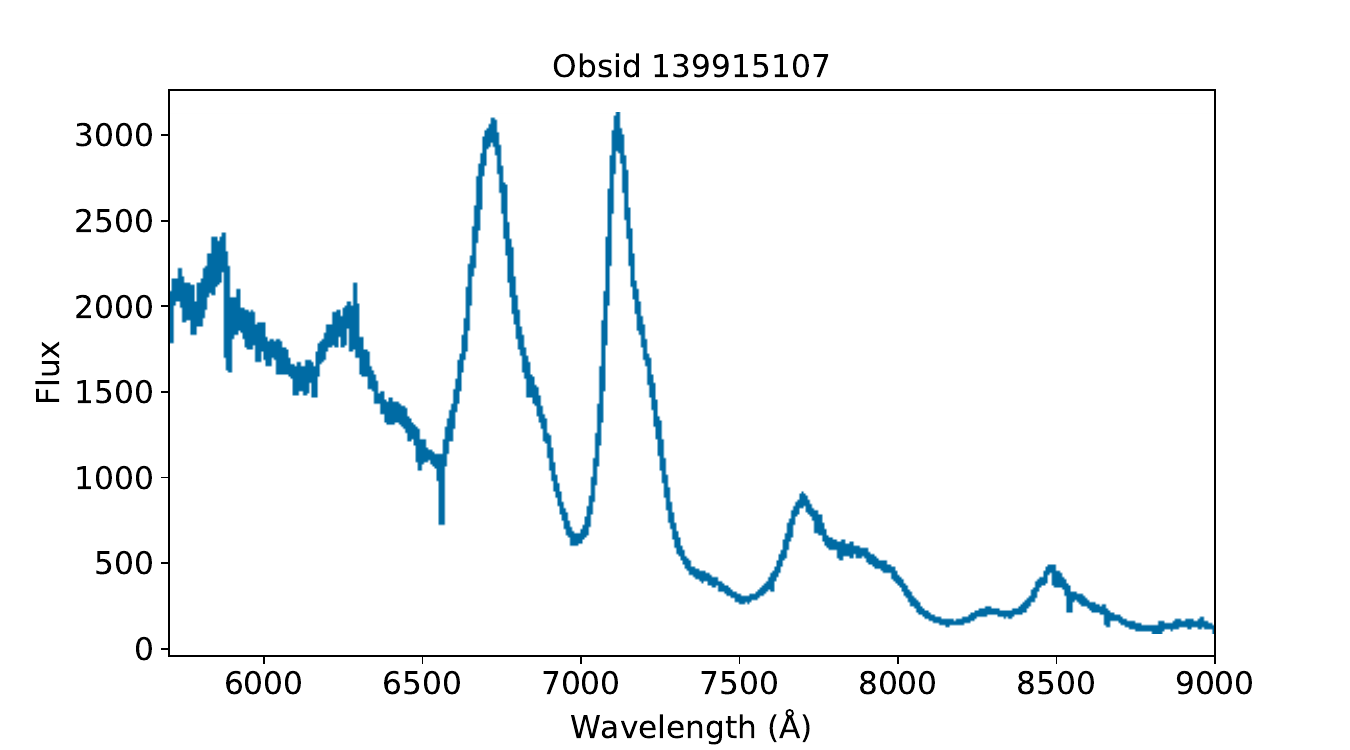}
\caption{Examples of the detected outliers in our sample. We show the LAMOST spectra in blue and the VAE reconstructions in orange. The top panel shows a spectrum with the strong H$\alpha$ and nebular forbidden emissions. The middle and bottom panels show the normalized H$\alpha$ region and the original red band (5700-9000 \AA) of a abnormal spectrum, respectively.}
\label{fig:outliers}
\end{figure}

In order to get reliable measurements of the chromospheric emissions, we need to firstly identify the abnormal spectra in our sample, which will influence our analysis severely. We calculated the RMSE of the residual spectra in the region outside the active lines to assess the quality of the model fitting and found the large miss-matched cases. The latent features generated by the encoder is also useful for the anomaly detection. The outliers in the latent space are too different from the training data sample. The most common types of the outliers in our sample are the strong emission lines, such as the cosmic rays and the nebular forbidden emission lines. The strong nebular emissions can lead to a very large residual EWs of the H$\alpha$ line. Two examples of the outliers, the strong nebular emissions around H$\alpha$ line and an abnormal spectrum, are shown in Fig. \ref{fig:outliers}.

One should note that the definition of the anomaly is relative. For instance, the nebular emissions worsen the analysis on the stellar magnetic activity but may interest some authors in other research field.

\subsection{Chromospheric emissions}

The H$\alpha$ and Ca~{\sc ii} IRT spectral lines are very sensitive to the chromospheric activity of cool stars, and thus are used as the stellar activity indicators. The resolution of the R1800 spectra is enough to analyse the spectra profiles, and thus we employed the spectral subtraction technique to investigate the stellar chromospheric activity in the K2 field. Using the VAE, we generated the inactive templates for all sample stars and derived the residual spectra. The excess EWs of H$\alpha$ line were derived from the integration of the residual flux over a bandwidth of 20 \AA\ centring at 6563 \AA. For Ca~{\sc ii} IRT lines, we integrated the residual flux over a bandwidth of 10 \AA\ around 8498 \AA, 8542 \AA\ and 8662 \AA.

\begin{figure*}
\centering
\includegraphics[width=0.96\textwidth]{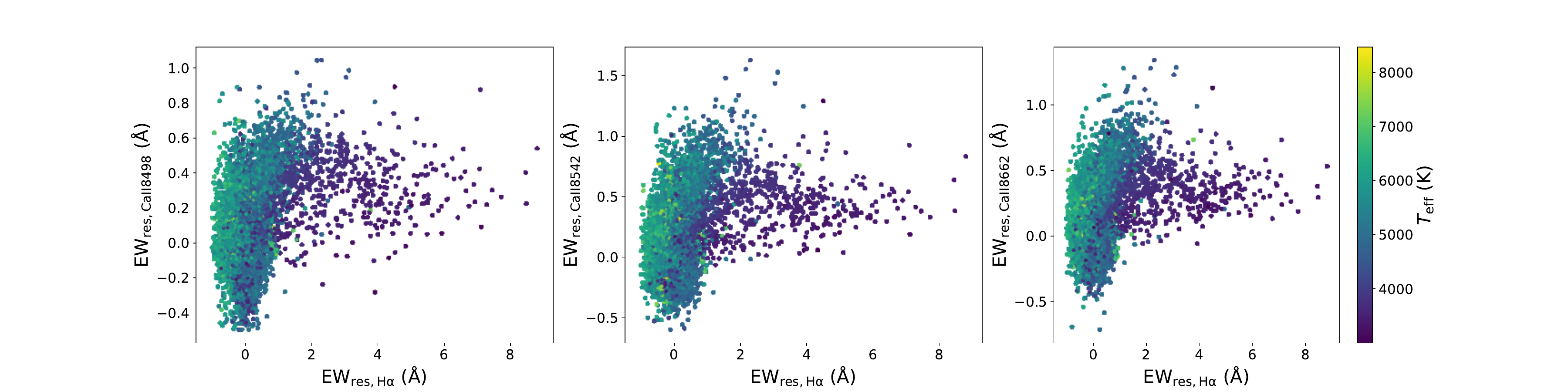}
\caption{The relationship pairs of the residual EWs of H$\alpha$ and Ca~{\sc ii} IRT. The $T_{\rm eff}$ is shown in colours.}
\label{fig:ews}
\end{figure*}

We show the relationship pairs between the residual EWs of H$\alpha$ and Ca~{\sc ii} IRT lines in Fig. \ref{fig:ews}, where the stellar $T_{\rm eff}$ derived by the LASP are shown in colour. The EWs of these lines of F- to K-type star shows a clear correlation, whereas those of M-type stars are more scattered. \citet{walkowicz2009} revealed that the less active M-type stars show weak correlation between the chromospheric emissions of H$\alpha$ and Ca~{\sc ii} K lines.

In this work, we set thresholds for the detection of the magnetically active stars in the K2 field as follows:

1. The residual EW of H$\alpha$ line is larger than 1 \AA.

2. The residual EWs of Ca~{\sc ii} IRT lines are larger than 0 \AA.

3. The RMSE of the residual spectrum outside the active lines is less than 0.02.

Note that these thresholds are arbitrary and are used to select the highly active stars with the residual emissions of H$\alpha$ and Ca~{\sc ii} IRT lines, and thus the stars showing lower level of the activity are omitted. To avoid other emission sources than the chromospheric activity, we also excluded spectra of emission stars hotter than 6000 K. As a result, we find 1176 spectra of 799 active stars. 633 spectra with the residual H$\alpha$ emissions were excluded from the active sample, because of no residual emission in Ca~{\sc ii} IRT lines or large RMSE. The H$\alpha$ and Ca~{\sc ii} IRT lines are among the most common indicators of the chromospheric activity of cool stars \citep{west2004,frasca2016}, but there are still other sources can cause the observed emission of the H$\alpha$ line, and the further identification of the activity requires more photometric and spectroscopic observations.

The chromospheric emission is highly dependent on the effective temperature, and thus the residual EWs are not suitable for the comparison between stars with different spectral types. We used the line flux ($F_{\rm H\alpha}$) and the ratio between the luminosity in H$\alpha$ line and the bolometric luminosity ($R'_{\rm H\alpha}=L_{\rm H\alpha}/L_{\rm bol}$) to quantify the emissions of H$\alpha$ line. $F_{\rm H\alpha}$ was calculated as the equation below:
\begin{equation}
F_{\rm H\alpha} = F_{6563}  EW_{\rm res,H\alpha}
\end{equation}
where F$_{6563}$ is the continuum flux at the H$\alpha$ line centre. We derived the continuum flux at 6563 \AA\ from the medium-resolution PHOENIX model stellar spectra \citep{husser2013} with the $T_{\rm eff}$ and the $\log g$. We used the equation in \citet{frasca2015} to estimate $R'_{\rm H\alpha}$, which is distance-independent, as below:
\begin{equation}
R'_{\rm H\alpha} = \frac{L_{\rm H\alpha}}{L_{\rm bol}} = \frac{F_{\rm H\alpha}}{\sigma \text{$T^{4}_{\rm eff}$}}
\end{equation}
where $\sigma$ is the Stefan–Boltzmann constant.

\begin{figure}
\centering
\includegraphics[width=0.45\textwidth]{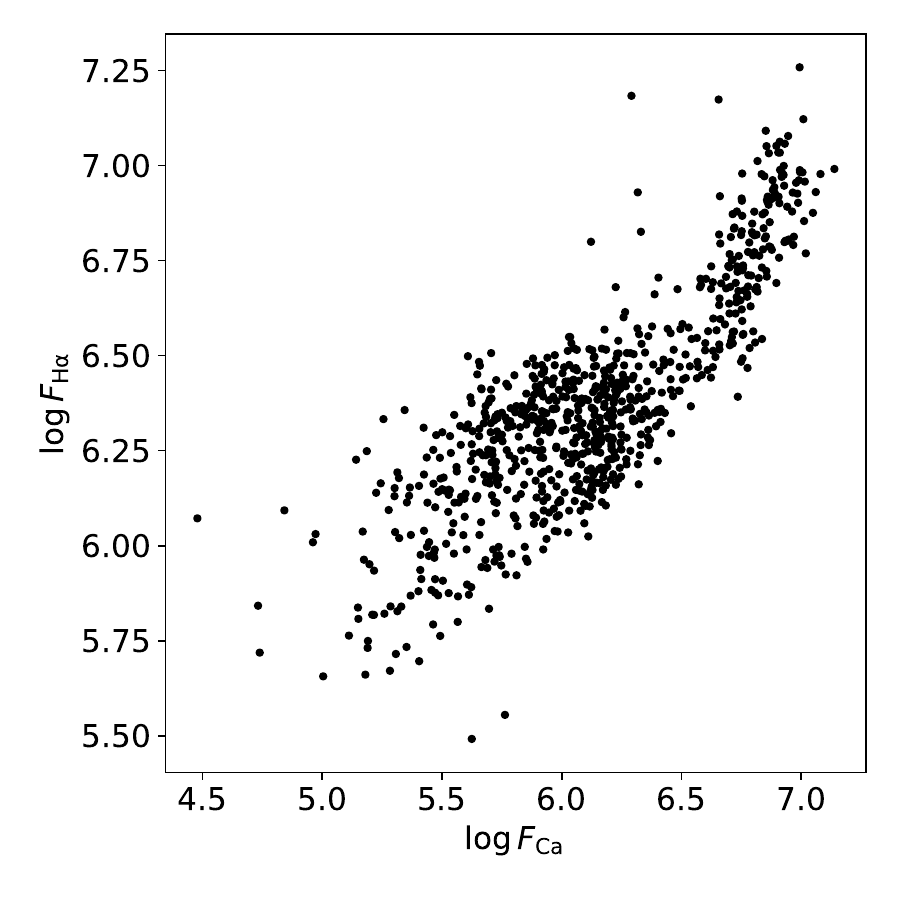}
\caption{$F_{\rm H\alpha}$ as a function of $F_{\rm Ca}$ for active stars.}
\label{fig:fhafca}
\end{figure}

We also calculated these quantities for Ca~{\sc ii} IRT, and used the sum of the values of Ca~{\sc ii} 8498, 8542 and 8662 lines to derive $F_{\rm Ca}$ and $R'_{\rm Ca}$. Fig. \ref{fig:fhafca} shows relationship between $F_{\rm Ca}$ and $F_{\rm H\alpha}$ of the active stars. The surface line fluxes of H$\alpha$ and Ca~{\sc ii} IRT lines are correlated well to each other. We list the quantities of the residual emissions of the chromospheric indicators for the detected active stars in Table \ref{tab:active}, which is only available online.

\begin{table*}
  \caption{The stellar parameters and the quantities of the residual emissions of the detected active stars in LK2 dataset. The full version is only available online.}
  \label{tab:active}
 \begin{tabular}{lcccccccccccc}
  \hline
  Obsid & EPIC & $T_{\rm eff}$ & $\log g$ & [Fe/H] & EW$_{\rm H\alpha}$  & EW$_{\rm Ca8498}$  & EW$_{\rm Ca8542}$  & EW$_{\rm Ca8662}$ & $\log F_{\rm H\alpha}$ &$\log F_{\rm Ca}$& $ \log R'_{\rm H\alpha}$ & $\log R'_{\rm Ca}$\\
        &      & K   & & dex   & \AA        & \AA          & \AA       & \AA   &        &   & &\\
  \hline
 40607226&   202073489&     5655&  4.08&   0.03&  1.06&  0.76&  1.07&  1.00&     6.85&     7.08&    -3.91&    -3.68\\
 41610221&   211657018&     4934&  2.68&  -0.56&  1.03&  0.12&  0.41&  0.37&     6.58&     6.38&    -3.95&    -4.15\\
 41614054&   211712465&     3948&  5.24&   0.00&  2.18&  0.62&  0.78&  0.67&     6.41&     6.38&    -3.73&    -3.76\\
 41704204&   201463400&     4859&  3.70&  -0.37&  1.25&  0.53&  0.77&  0.69&     6.64&     6.70&    -3.86&    -3.80\\
 45202125&   212116340&     5478&  4.29&  -0.24&  1.39&  0.42&  0.77&  0.77&     6.91&     6.87&    -3.80&    -3.83\\
 \hline
 \end{tabular}
\end{table*}

\begin{figure}
\centering
\includegraphics[width=0.4\textwidth]{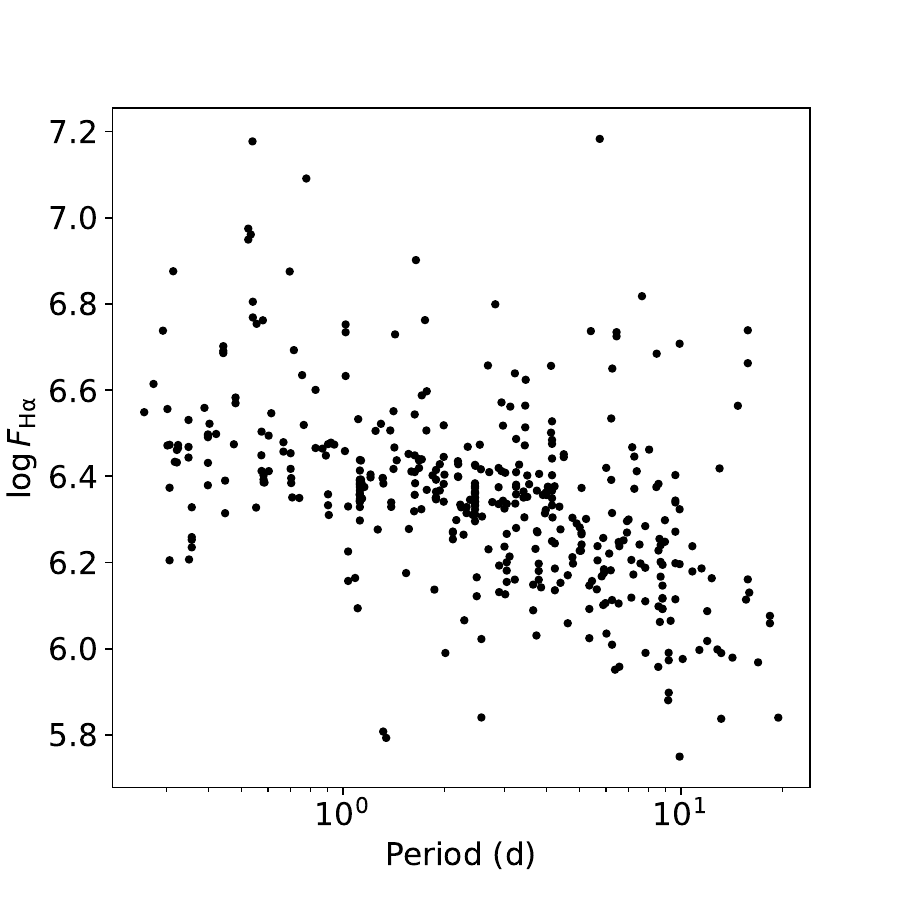}
\includegraphics[width=0.4\textwidth]{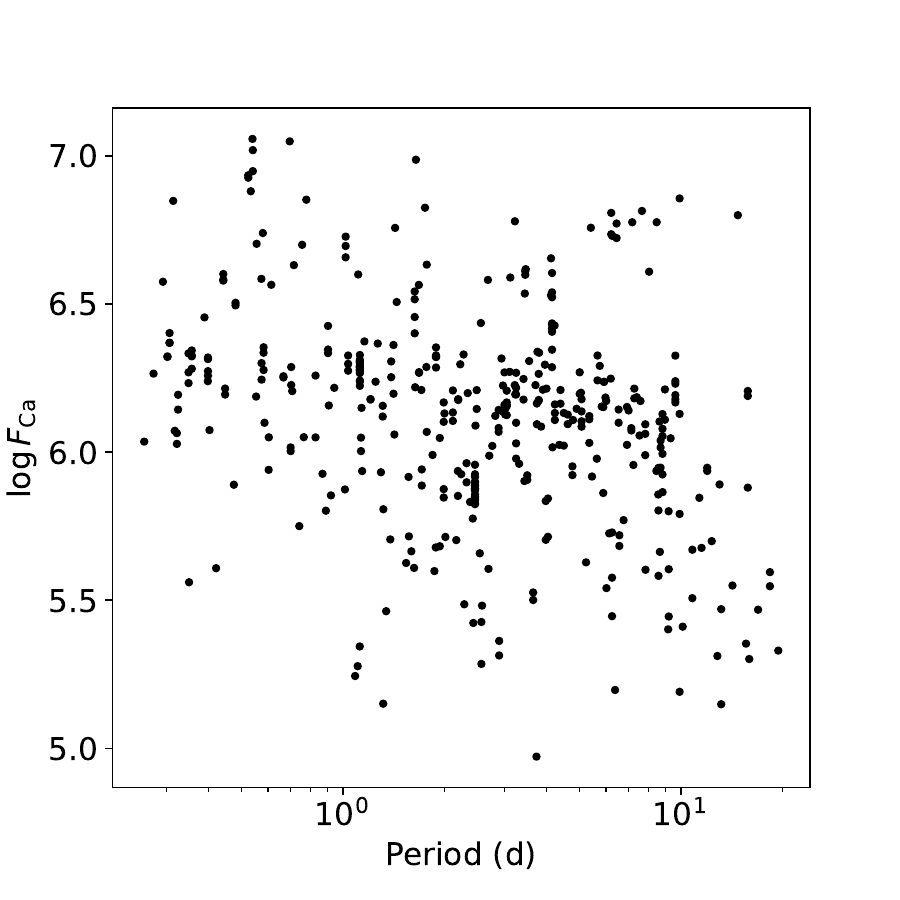}
\caption{$F_{\rm H\alpha}$ (top) and $F_{\rm Ca}$ (bottom) as functions of the rotational period.}
\label{fig:fprot}
\end{figure}

\begin{figure}
\centering
\includegraphics[width=0.4\textwidth]{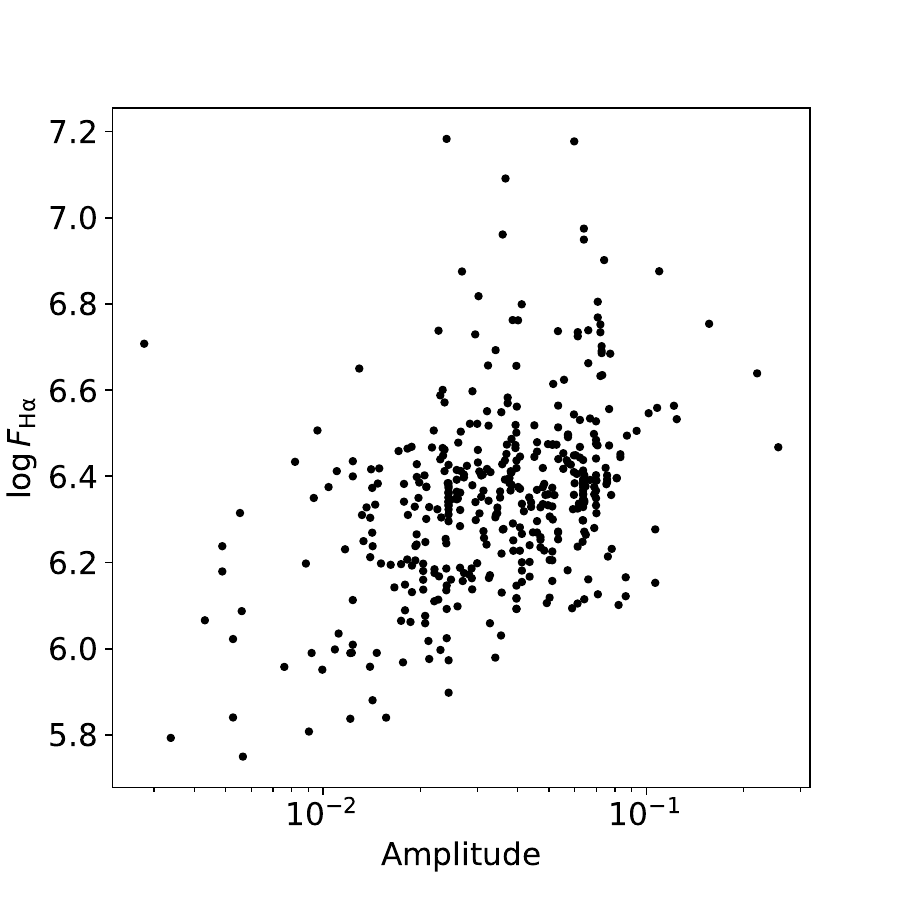}
\includegraphics[width=0.4\textwidth]{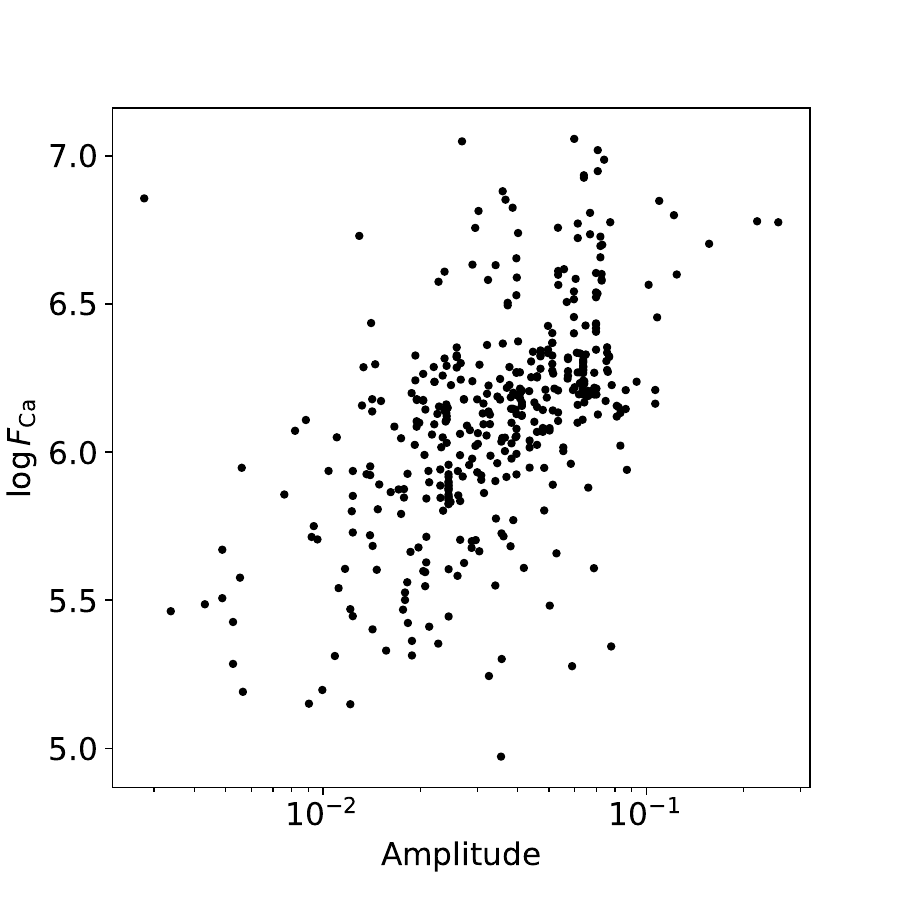}
\caption{Same as Fig. \ref{fig:fprot}, but for the amplitude.}
\label{fig:famp}
\end{figure}

The \textit{Kepler} satellite has obtained a large set of the high-precision stellar light curves. Vast authors have conducted studies on the variable stars in the K2 field using the photometry, and fruitful results were presented recently. We adopted the dominant Lomb-Scargle period and the amplitude of the binned phase curve derived by \citet{armstrong2016} to reveal how the excess emissions in H$\alpha$ and Ca~{\sc ii} IRT lines in our sample correlate to these light curve parameters. \citet{armstrong2016} has classified the types of the variable stars in the K2 field with the K2 light curves using the machine learning algorithm. We thus removed the star that shows variable flux due to the reason other than magnetic activity, such as eclipsing, from the statistics. The results are shown in Fig. \ref{fig:fprot} and \ref{fig:famp}, and we also show examples of the K2 light curves and the corresponding R1800 spectra of the active stars in Fig. \ref{fig:lc}. Both emissions of H$\alpha$ and Ca~{\sc ii} IRT lines increase with decreasing rotational period. While, the correlation between the emissions and the amplitude of phase curve is clearer, that is, the chromospheric emissions increase with increasing amplitude. The figures reveal that the rapidly rotating stars show a high level of magnetic activity, which is in good agreement with the prediction of the $\alpha\Omega$ dynamo. The stellar rotation is one of the most important ingredients for generating the magnetic field of stars. The activity-rotation relationship has been widely investigated by many authors with various indicators, from X-ray to infrared \citep{pallavicini1981,reiners2014,farrish2021}. The increase trend of the emissions of the chromospheric lines with increasing amplitude of the light curves indicates a clear positive correlation between the photospheric and chromospheric activity, since the photometric flux variations of these stars are mainly attributed to the spot modulation \citep{armstrong2016}. The photosphere and chromosphere relations were reported by \citet{radick1998} and \citet{zhangj2020}. \citet{berdyugina1999} revealed that the position of the chromospheric active region is related to the starspot activity on \mbox{II Peg} using high-resolution spectra.

\begin{figure}
\centering
\includegraphics[width=0.45\textwidth]{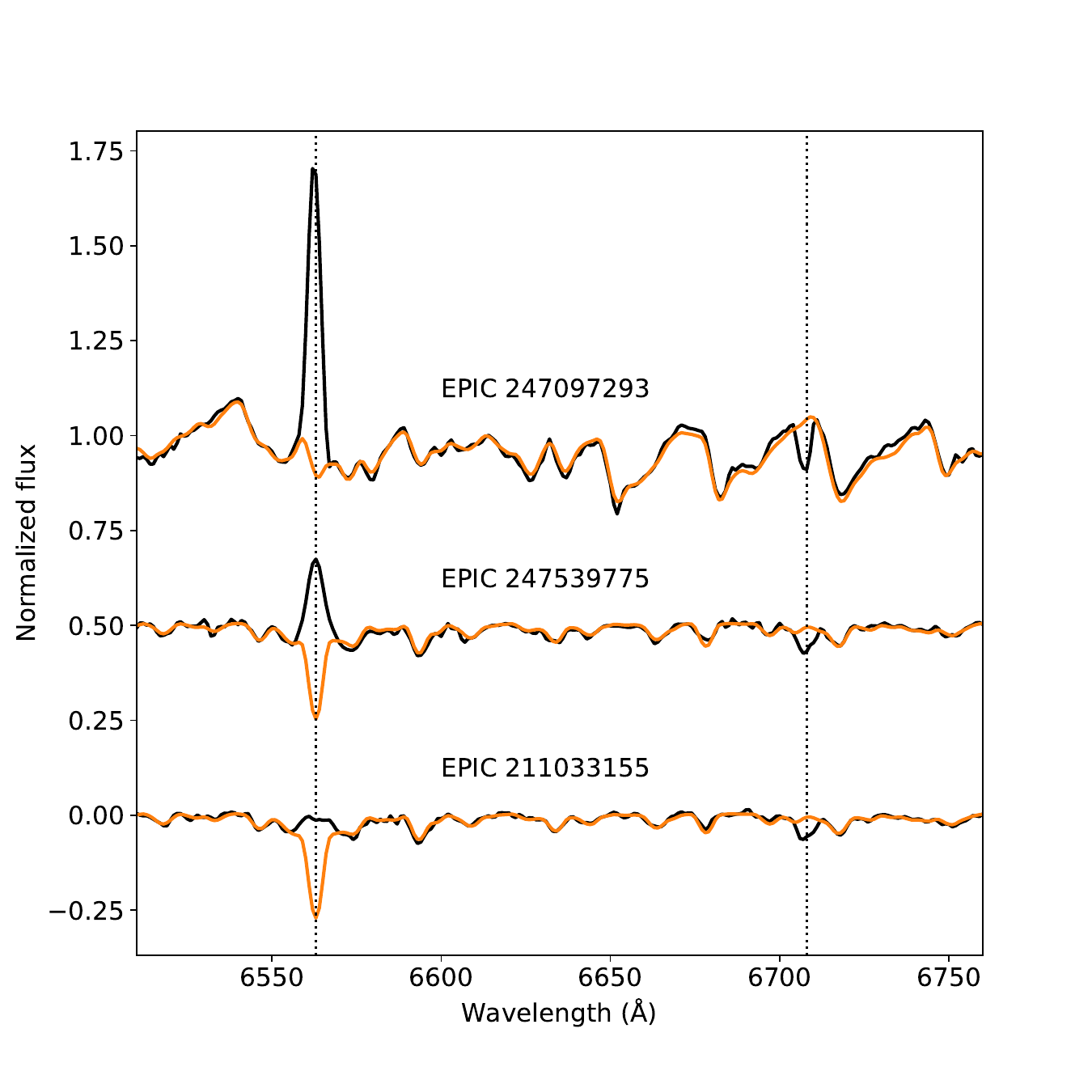}
\caption{Examples of the chromospherically active stars with the excess absorption at 6708 \AA, which indicates a high lithium abundance. The LAMOST spectra are shown in black and the VAE reconstructions are in orange.}
\label{fig:li}
\end{figure}

In our dataset, the R1800 spectra also cover lithium doublet at 6708 \AA. In addition to the H$\alpha$ emissions, we also detected the stars with excess absorptions in the region of 6708 \AA, which means a high lithium abundance. For instance, three spectra of the magnetically active stars with excess absorption of lithium are shown in Fig. \ref{fig:li}. Some of them are the Taurus members derived by \citet{rebull2020}, which means that these are active young stars. Pre-main-sequence stars show a high level of the magnetic activity due to the relatively high rotational rate and they have a high lithium abundance due to the young age. Lithium is an important element, whose abundance reflects the evolutionary stage of stars. Some F- and K-giants show high lithium abundance \citep{brown1989,lyubimkov2012}, which challenges the stellar evolution knowledge. Many authors identified the Li-rich stars in LAMOST low-resolution and medium-resolution spectroscopic surveys (e.g. \citealt{gao2021,wheeler2021}). The VAE model learns from the majority of the dataset and thus can detect rare targets like lithium-rich stars, in an unsupervised manner.

\subsection{Chromospheric activity in R250 spectra}

Although the spectral resolution is very low, the slitless spectroscopic survey can be used to search for the interesting celestial objects (e.g. \citealt{worseck2008,martayan2010,ohyama2018}). In order to explore the possibility of detecting and characterizing the magnetically active stars with the planned CSST slitless spectroscopic survey, we analysed the simulated R250 spectra and the corresponding VAE reconstructions.

Due to the low resolution of the R250 spectra, we focus on the emission of H$\alpha$ line, which affects a wide wavelength range. We measured the EWs of the H$\alpha$ line through integrating over a wider bandwidth of 25\AA\  (6550--6575 \AA), as described by the equation below:
\begin{equation}
EW = \int \frac{F_{\lambda}-F_{c}}{F_{c}} d\lambda
\label{eq:ew}
\end{equation}
where we derived the continuum flux $F_{c}$ by averaging the fluxes of two 50 \AA\ bands at two sides of H$\alpha$ line (6500-6550 \AA\ and 6575-6625 \AA), same as that used by \citet{west2004}.

\begin{figure}
\centering
\includegraphics[width=0.45\textwidth]{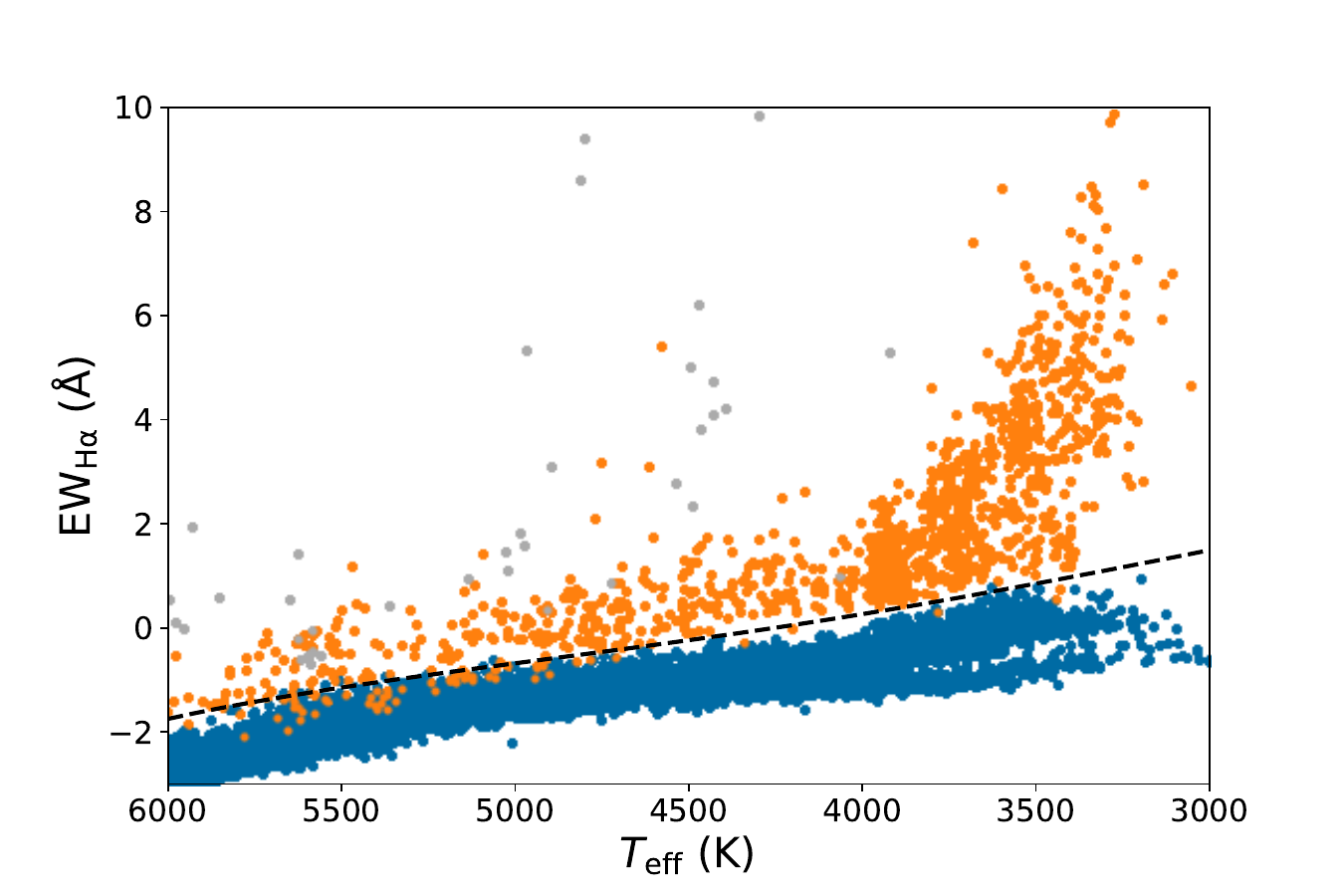}
\includegraphics[width=0.4\textwidth]{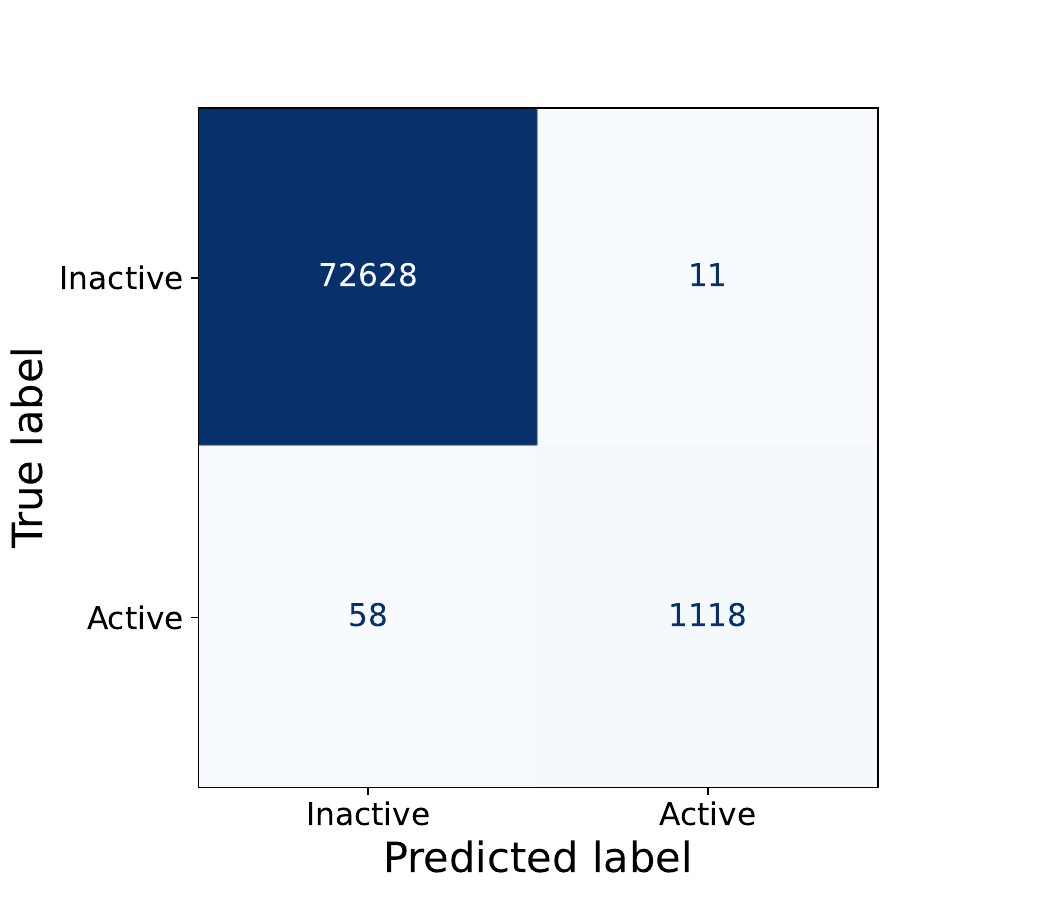}
\caption{The upper panel shows the EWs of H$\alpha$ derived from R250 spectra as a function of the $T_{\rm eff}$. The blue, orange and grey dots denote the inactive, active and nebular emission stars, respectively. The dashed line is the boundary between inactive and active stars derived with the SVM model. The lower panel shows the confusion matrix of the classification on the active and inactive stars cooler than 6000 K.}
\label{fig:hateff}
\end{figure}

In Fig. \ref{fig:hateff}, we show the EWs of H$\alpha$ of selected stars as a function of the $T_{\rm eff}$ derived by the LASP. The red points denotes the active stars detected from the R1800 dataset and the blue points are inactive stars, whose EWs derived from the R1800 spectra are less than 0.5 \AA. The grey dots are the stars with strong nebular emissions. The forking structure of EWs of inactive stars at the low $T_{\rm eff}$ of 3000--4000 K is due to the different $\log g$. The result shows that we can use the EWs of H$\alpha$ line of the inactive stars, which are detected in the R1800 dataset, as the basal values to search for the magnetically active stars with the EWs higher than the basal one at their $T_{\rm eff}$. We tried to separate the active and inactive stars cooler than 6000 K with a support vector machine (SVM) model. The EWs of the H$\alpha$ line and the $T_{\rm eff}$ of stars were used as the features. We fitted the SVM model with a polynomial kernel to 1176 active and 72639 inactive samples, and the boundary given by the SVM model is shown as the dashed line in Fig. \ref{fig:hateff}. The EWs between the detected active and inactive stars were separated more clearly with decreasing stellar $T_{\rm eff}$. We also display the confusion matrix of the classification in Fig. \ref{fig:hateff} to show the performance of the SVM model. As can be seen from the figure, the model can identify 95 per cent of the active stars. However, it is impossible to distinguish the chromospheric emission and the nebular emission, due to the very low spectral resolution. We additionally plot the calculated EWs of the H$\alpha$ line for the spectra with the nebular emissions in Fig. \ref{fig:hateff}. The nebular emissions can lead to a false positive detection on the stellar magnetic activity with the R250 spectra.

\begin{figure*}
\centering
\includegraphics[width=0.8\textwidth]{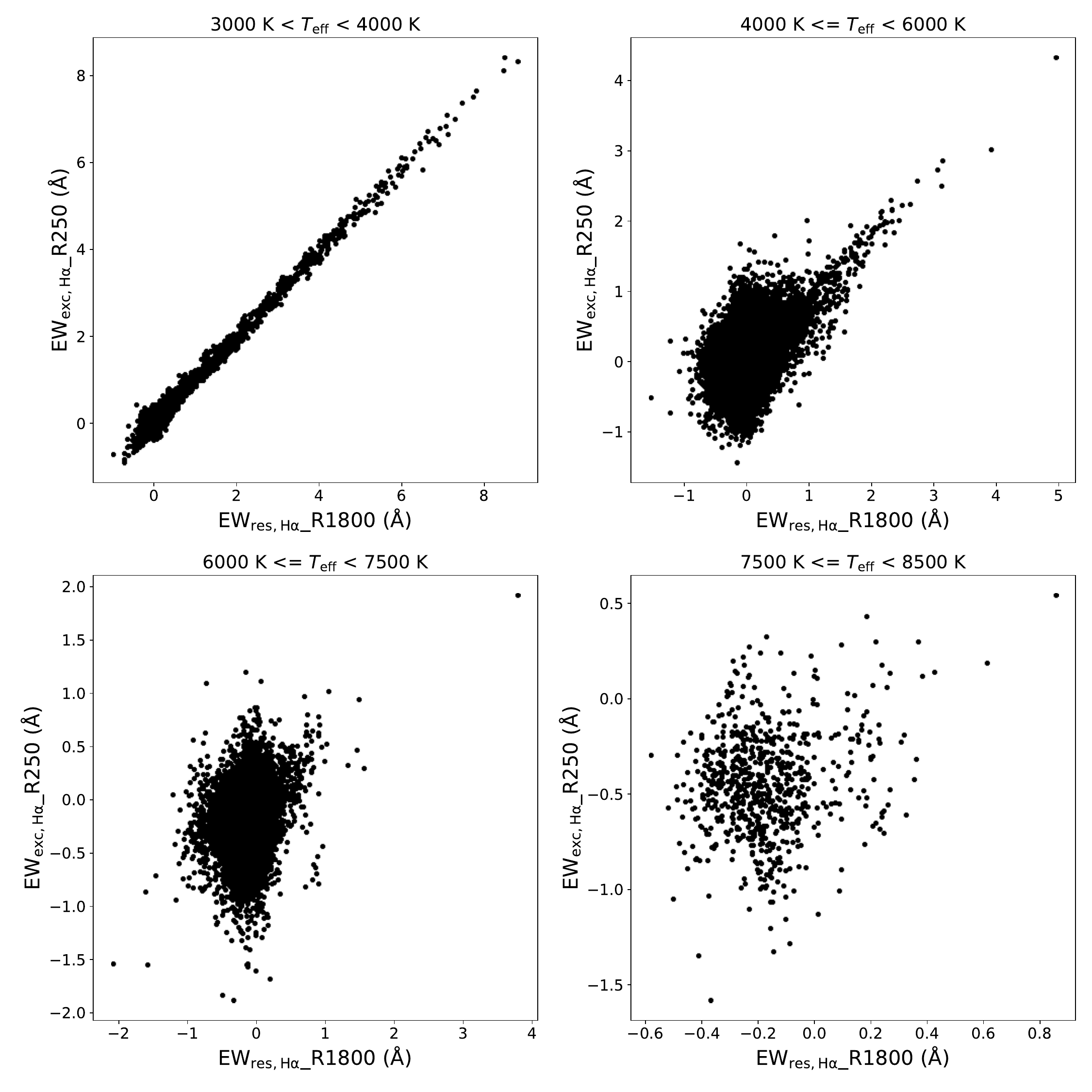}
\caption{The excess EWs of H$\alpha$ line measured from R250 spectra as a function of those from R1800 spectra for the different effective temperature ranges.}
\label{fig:ewew}
\end{figure*}

The above method requires the estimations of stellar $T_{\rm eff}$ and it does not take account of other parameters like the $\log g$. The simultaneous multi-band photometry is helpful for the determination of the spectral type of stars. Using the VAE modelling, We can also derive the excess EWs of H$\alpha$ without stellar parameters. Different from how we estimated those for the R1800 spectra, we respectively calculated the EWs of H$\alpha$ line in the R250 spectra and the corresponding VAE reconstructions, and then computed the difference between them as the excess EWs of H$\alpha$ line (EW$_{\rm exc,H\alpha}$). We display the correlation between the excess EWs of H$\alpha$ line derived from the R1800 and R250 spectra in Fig. \ref{fig:ewew}. Note that the values of EWs larger than 10 are removed, in order to display the correlations more clearly. The EWs derived from the spectra with different resolutions are correlated well to each other, for the stars cooler than 6000 K, though the absolute value is highly dependent on the sampling rate of the spectrum. It means that we can use the EWs derived from the low-resolution slitless spectra to roughly characterize the magnetic activity of cool stars. Meanwhile, VAE can also be used to detect anomaly in the slitless spectra.

\section{Conclusions}

The big data era in astronomy is coming. Recent spectroscopic surveys have obtained tens of millions of stellar spectra. Combined with the photometric survey data, like Kepler/K2 photometry, we can go much deeper into the mystery of the stellar magnetic field. With the large FoV and deeper limiting magnitude, the planned CSST all-sky slitless spectroscopic survey will deliver an unprecedented large database of stellar spectra, especially for the faint low-mass stars. Fast and efficient tools for analysing the large-scale data of the low-resolution spectroscopic surveys are urgently required.

In this work, we obtain 132086 low-resolution stellar spectra from the LK2 database and use the VAE to model them to investigate the stellar magnetic activity of stars in the K2 field. We also simulate the CSST slitless spectra by degrading these LAMOST spectra, to explore the ability of detecting and characterizing the magnetically active stars in the future CSST all-sky survey.

The two wavelength ranges of 6370-6770 \AA\ and 8380--8780 \AA\ of the LAMOST spectra are extracted and normalized to construct the R1800 dataset. Meanwhile, the degraded spectra in the wavelength range of 6000--7500 \AA\ are used to form the R250 dataset. We build the VAE model which extracts the features of the low-resolutions spectra and generates the corresponding inactive template spectra, in an unsupervised way. We separately train two VAE models, with the same architecture of hidden layers, on the R1800 and the simulated R250 spectra. Both of them can efficiently synthesize the inactive template spectra, which demonstrates that the architecture of the VAE model is general to handle different kinds of spectral data with a proper training. 

We detect abnormal spectra in our sample using the RMSE of the residual spectra and the latent distribution generated by the encoder of the VAE. The abnormal spectra are mainly due to the cosmic-rays and strong nebular emissions. These features can severely affect the analysis on the stellar magnetic activity, especially for the nebular forbidden emission lines on both sides of H$\alpha$ line.

We measure the excess emissions of the chromospheric indicators, H$\alpha$ and Ca~{\sc ii} IRT, from the residual spectra to detect the magnetically active stars in the K2 field. The excess emissions of H$\alpha$ are correlated well to those of Ca~{\sc ii} IRT. We find that the strength of the chromospheric emissions of variable stars is clearly correlated to the rotational period and the light curve amplitude.

Additionally, we demonstrate the ability of the VAE model to detect strong absorption in lithium 6708 \AA. Some of active stars in the K2 field with lithium excess absorption are young stars, in our sample.

Due to the very low resolution of R250 spectra, we derive EW of H$\alpha$ line by integrating over a wider bandwidth. Using the active and inactive stars detected in the R1800 dataset, We find that the EWs of H$\alpha$ line can be used for the detection of active stars. However, it is difficult to distinguish the chromospheric emission of H$\alpha$ line and the strong nebular emissions in R250 spectra. Using the VAE, the excess EWs derived from R250 spectra are in good agreement with values derived from R1800 spectra for cool stars. We thus can characterize the magnetic activity of cool stars with the slitless spectra using a unsupervised model.

The planned CSST all-sky survey will conduct the simultaneous multi-band photometry and slitless spectroscopy, which is very suited for the study of stellar magnetic activity. Due to the deeper limiting magnitude and a large FoV, CSST survey will deliver an unprecedented big data of the spectra and photometry of vast stars. It is critical to develop methods to detect peculiar objects in the future CSST all-sky survey. The selected interesting targets are also suitable for follow-up observations by the ground-based large-aperture telescopes.

The CSST slitless spectrum will have a higher SNR and no telluric line contaminations, compared to the LAMOST spectroscopic observations. Moreover, a large number of stars will be imaged by the CCD of the CSST in the same exposure, which benefits the accurate flux calibration. Whereas, the flux calibration for the LAMOST spectra is relative and is much dependent on the stellar atmosphere model \citep{song2012}.

We also note that the recent simulation for CSST observational data is not sufficient, and the multi-band photometry and slitless spectroscopy simulations for the CSST all-sky survey is still in progress. Later, we will take advantage of those simulations to investigate more aspects related to stellar magnetic activity.

After the launch of the CSST, a simple scenario is that we use the same samples covered by the recent spectroscopic survey (e.g. SDSS, LAMOST) to train the machine learning model and calibrate the quantities of the activity indicators. A well-trained model will be very efficient for the detection and analysis on large-scale sample of the magnetically active cool stars. Combined with the machine learning model, we will develop a pipeline for the measurement of the stellar magnetic activity in the future.

\section*{Acknowledgements}
We are grateful to the anonymous referee for valuable comments and suggestions that significantly improved this paper. We acknowledge the science research grant from the China Manned Space Project with NO. CMS-CSST-2021-B07. This study is supported by the National Natural Science Foundation of China under grants Nos.10373023, 10773027, U1531121, 11603068 and 11903074. Guoshoujing Telescope (the Large Sky Area Multi-Object Fiber Spectroscopic Telescope LAMOST) is a National Major Scientific Project built by the Chinese Academy of Sciences. Funding for the project has been provided by the National Development and Reform Commission. LAMOST is operated and managed by the National Astronomical Observatories, Chinese Academy of Sciences. This paper includes data collected by the \textit{Kepler} mission. Funding for the \textit{Kepler} mission is provided by the NASA Science Mission directorate. 

\section*{Data availability}
The raw LAMOST spectra can be obtained at the official website \url{http://www.lamost.org}. We provide the code of the VAE and the model weights at \url{https://github.com/xylib/vae-for-spectroscopic-survey}.

\bibliographystyle{mnras}
\bibliography{csst}

\appendix

\section{Light curves and spectra of the active stars}

\begin{figure*}
\centering
\includegraphics[width=0.95\textwidth]{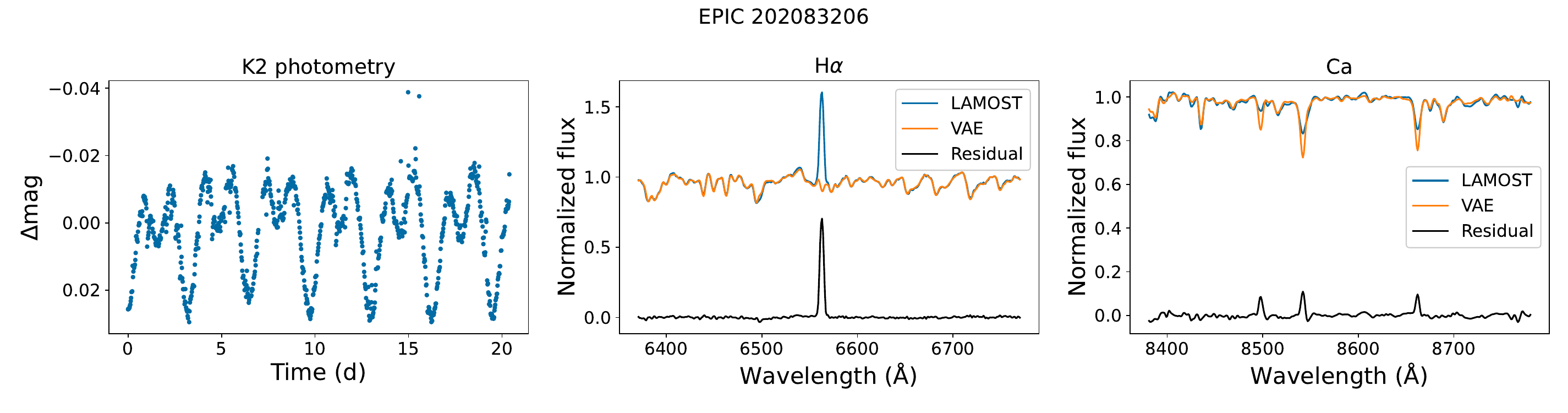}
\includegraphics[width=0.95\textwidth]{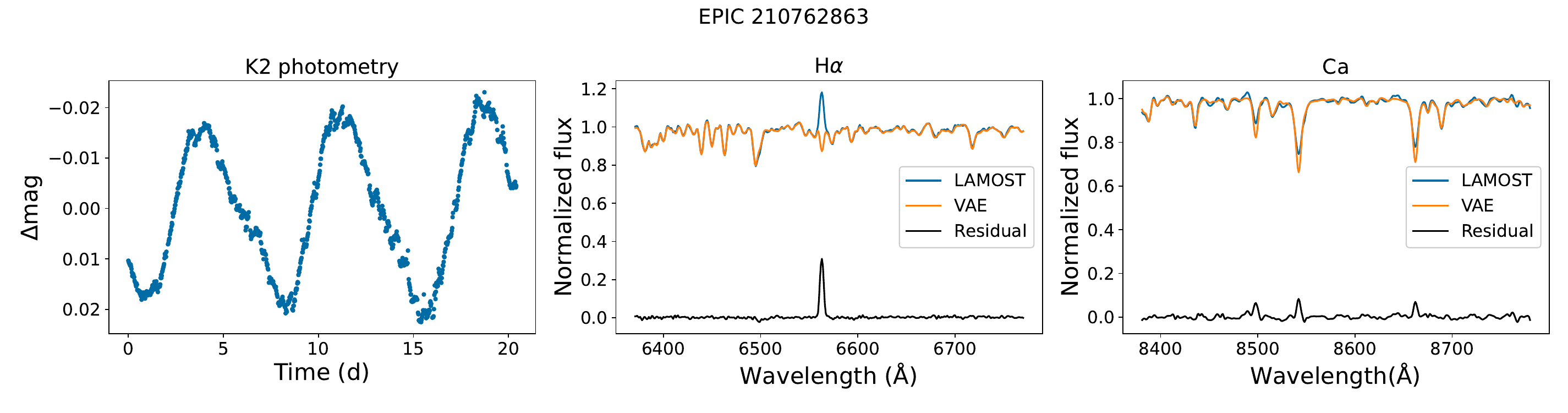}
\includegraphics[width=0.95\textwidth]{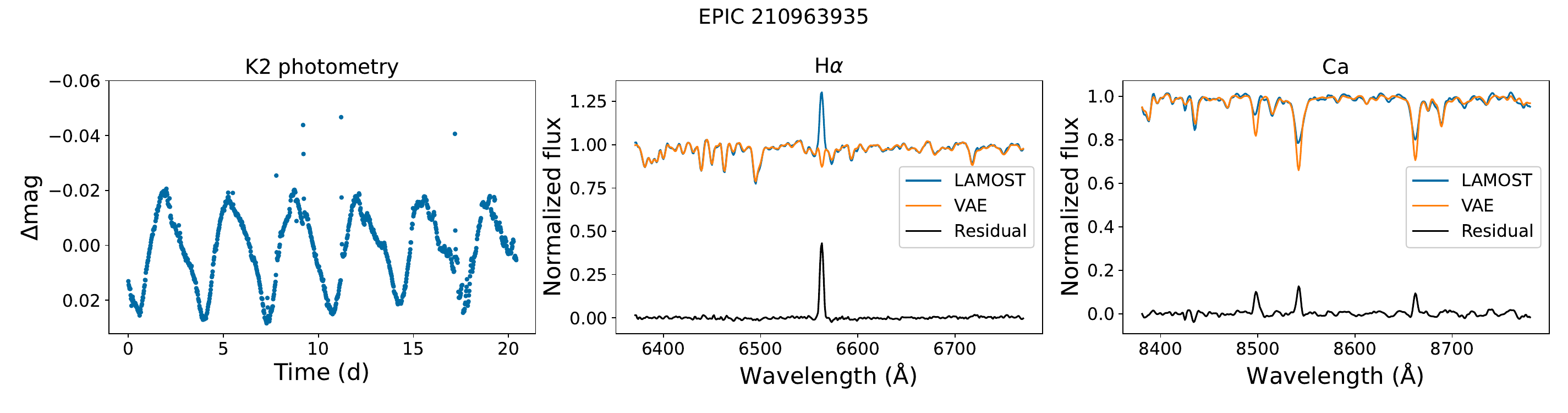}
\includegraphics[width=0.95\textwidth]{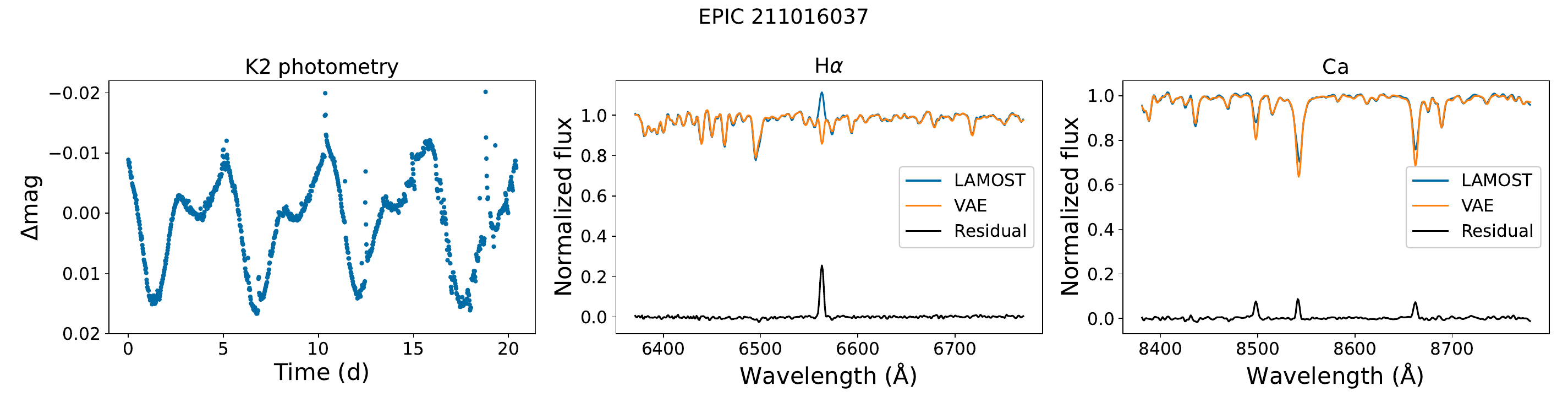}
\includegraphics[width=0.95\textwidth]{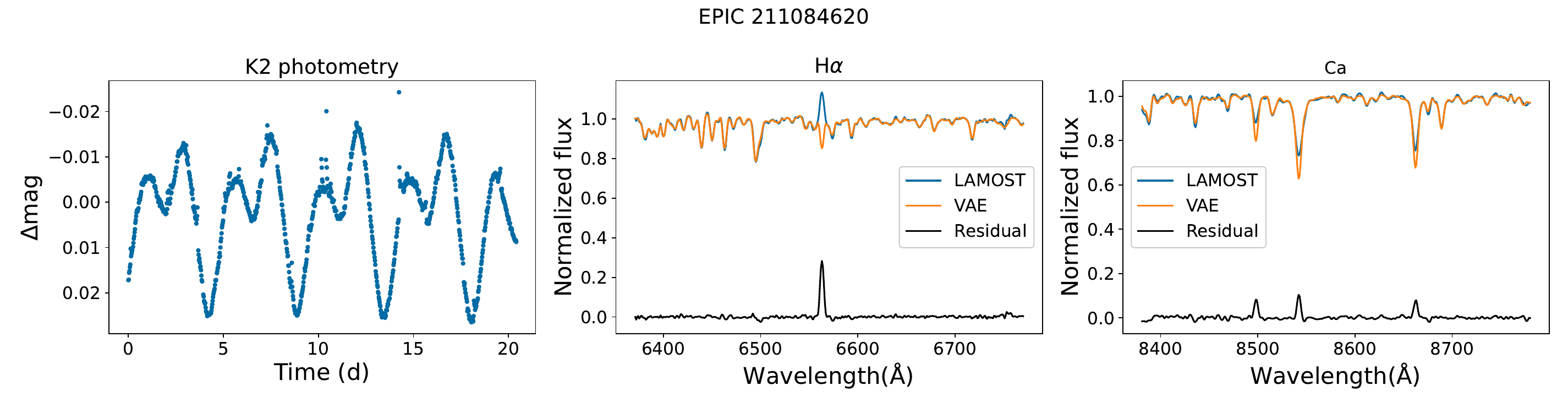}
\caption{Examples of the K2 light curves and the corresponding R1800 spectra of the active stars.}
\label{fig:lc}
\end{figure*}

In Section 4.3, we compare the period and amplitude of the K2 light curves of the active stars derived by \citet{armstrong2016} and the residual emissions of the chromospheric indicators in the R1800 spectra. We display some examples of the K2 light curves and the corresponding R1800 spectra of the active stars in Fig. \ref{fig:lc}. As can be seen from this figure, these magnetically active stars show the residual emissions of the H$\alpha$ and Ca~{\sc ii} IRT lines as well as the rotational modulation of brightness probably caused by starspots on the stellar surface, and some of them also exhibit flare activity.

\label{lastpage}
\end{document}